\documentclass{aa}

\usepackage[tbtags,fleqn]{amsmath}
\usepackage{amsfonts}
\usepackage{pstricks, pst-plot}
\usepackage{graphicx}
\usepackage{natbib}

\newcommand{\e}{\mathrm e}
\newcommand{\diff}{\mathrm d}

\newcommand{\R}{\mathbb R}
\newcommand{\mincir}{\raise
  -2.truept\hbox{\rlap{\hbox{$\sim$}}\raise5.truept \hbox{$<$}\ }}
\newcommand{\magcir}{\raise
  -2.truept\hbox{\rlap{\hbox{$\sim$}}\raise5.truept \hbox{$>$}\ }}
\newcommand{\Log}{\log_{10}}


\begin{document}

\title{Optimal column density measurements from multiband
  near-infrared observations}
\author{M. Lombardi\inst{1,2}}
\offprints{M. Lombardi} 
\mail{mlombard@eso.org}
\institute{%
  European Southern Observatory, Karl-Schwarzschild-Stra\ss e 2,
  D-85748 Garching bei M\"unchen, Germany \and
  University of Milan, Department of Physics, via Celoria 16, I-20133
  Milan, Italy}
\date{Received ***date***; Accepted ***date***}
\abstract{%
  We consider from a general point of view the problem of determining
  the extinction in dense molecular clouds.  We use a rigorous
  statistical approach to characterize the properties of the most
  widely used optical and infrared techniques, namely the star count
  and the color excess methods.  We propose a new maximum-likelihood
  method that takes advantage of both star counts and star colors to
  provide an optimal estimate of the extinction.  Detailed numerical
  simulations show that our method performs optimally under a wide
  range of conditions and, in particular, is significantly superior to
  the standard techniques for clouds with high column-densities and
  affected by contamination by foreground stars.  \keywords{dust,
    extinction -- methods: statistical -- ISM: clouds -- infrared: ISM
    -- stars: formation} }

\maketitle

\section{Introduction}
\label{sec:introduction}

A detailed comprehension of the structure and physical properties of
dark molecular clouds is a critical step to understand fundamental
processes such as star and planet formation.  Still, after several
decades of investigations, very little is known about dark clouds,
about their relationship to the diffuse interstellar medium, and about
the composition and evolution of the dust grains present in the
clouds.

Molecular clouds are composed of approximately $99\%$ molecular
hydrogen and helium, but due to the absence of a dipole moment these
molecules are virtually undetectable at the low temperature ($\sim 10
\mbox{ K}$) that characterize these objects.  As a result, astronomers
have been using different tracers to study the density distribution of
molecular clouds.  Historically, the first technique was based on a
statistical analysis of the angular density of stars observed through
a cloud \citep{1923AN....219..109W, 1937dss..book.....B}.  This
method, known as star counts, uses the dust (which is responsible for
the extinction of light) as a tracer of $\mathrm{H}_2$ and
$\mathrm{He}$.

More recently, radio observations of carbon monoxide (CO) and its
isotopes have been used to study the spatial distribution and the
physical conditions of molecular clouds, under the assumption that CO
is a reliable tracer of the matter in the cloud, i.e.\ that the ratio
$N(\mathrm{CO}) / \bigl( N(\mathrm{H}^2) + N(\mathrm{He}) \bigr)$ is
approximately constant.  Radio spectroscopy techniques have provided
extremely effective in studying the most dense regions (above $10^{22}
\mathrm{ protons\ cm}^{-2}$) and, more importantly, give dynamical
information on the cloud structure.  However, with the advent of
various dust detectors in the near-infrared (NIR), far infrared,
millimeter, and sub-millimeter bands, it has become clear that several
poorly constrained processes (e.g., deviations from thermodynamic
equilibrium, opacity variations, chemical evolution, and more
importantly depletion of molecules) can significantly affect the
results of analyses based on radio observations
\citep[e.g.][]{1994ApJ...429..694L, 1999ApJ...515..265A,
  2004A&A...421.1087H}.

The reddening of background stars offers a natural method to study the
distribution of dust in molecular clouds, and thus the hydrogen column
density.  This technique can be better applied to the infrared bands,
which compared to optical bands are less affected by extinction and
are less sensitive to the physical properties of the dust grains
\citep{1990ARA&A..28...37M}.  Before the advent of large format array
cameras, the lack of instrumental sensitivity clogged infrared
observations to small, dense clouds
\citep[e.g.][]{1980ApJ...242..132J, 1982ApJ...262..590F,
  1984ApJ...282..675J, 1986MNRAS.223..341C}.  More recently, infrared
arrays have made it possible to measure thousands of stars and to
extend the original technique to entire molecular cloud complexes
\citep{1994ApJ...429..694L, 1999ApJ...512..250L, 2001Natur.409..159A,
  2001A&A...377.1023L}.  Such measurements are free of the
complications that plague molecular-line data and permit a detailed
analysis of the cloud density distribution.

Although the success of the NIR color excess method is evident, there
is still a need for a deeper understanding of its limitations,
statistical biases, and uncertainties.  The aim of this paper is
twofold: (i) to study in detail the statistical properties of the star
count and color excess methods and (ii) to describe a new, optimal
method based on a maximum-likelihood analysis.  The method is
described here for NIR observations, but could equally well be applied
to other infrared bands for which the extinction law and the intrinsic
color distribution are known with good accuracy.  The paper is
organized as follows.  In Sect.~\ref{sec:basic-relations} we introduce
our formalism and consider from a statistical point of view the
extinction of stars by a foreground cloud.  The two main techniques
used to obtain extinction measurements, namely the star count and the
NIR color excess methods, are discussed in
Sect.~\ref{sec:extinct-meas}.  In Sect.~\ref{sec:likelihood-approach}
we describe the new maximum-likelihood method and show by means of
numerical simulations that it performs excellently even in the
presence of a significant contamination by foreground stars.  Our
conclusions are briefly reported in Sect.~\ref{sec:discussion}.
Finally, in Appendix~\ref{sec:medi-relat-estim} we consider the
statistical properties of the median and of related estimators used
often in infrared studies of dark clouds to remove the effects of
foreground stars.

\section{Basic relations}
\label{sec:basic-relations}

Suppose that $N$ stars with magnitudes (on a given band) $\{ \hat m_n
\}$ are observed in a given region of the sky (hereafter we will use
the hat $\hat\ $ to denote measured quantities).  These observed
magnitudes depend on the original, unreddened, absolute star
magnitudes $\{ M_n \}$, on the star distances $\{ D_n \}$, on the
individual extinctions $\{ A_n \}$, and on the photometric errors $\{
\epsilon_n \}$.  In particular, we have
\begin{equation}
  \label{eq:1}
  \hat m_n = M_n + 5 \Log D_n - 5 + A_n + \epsilon_n = m_n +
  \epsilon_n \; ,
\end{equation}
where the distances $\{ D_n \}$ are taken to be expressed in parsecs.
The quantity $m_n$ that appears in the r.h.s.\ of this equation is the
reddened magnitude of the $n$-th star free from measurement errors
(i.e., the magnitude that would be observed in the limit of an
extremely long exposure time).

\subsection{Single-band probability distributions}
\label{sec:prob-distr}

In order to consider the most general situation in a proper
statistical way, we introduce several probability distributions:
\begin{description}
\item[$\rho(M, D)$:] the probability distribution (density) of stars
  with absolute magnitude $M$ at distance $D$;
\item[$p_A(A | D)$:] the probability distribution of extinction for
  objects located at distance $D$, i.e.\ the probability of an
  extinction of $A$ magnitudes on a star given that its distance is
  $D$;
\item[$p_\epsilon(\hat m | m)$:] the probability distribution of
  photometric measurement errors, i.e.\ the probability of measuring a
  magnitude $\hat m$ for a star given that its true reddened magnitude
  is $m$.
\end{description}
These probability distributions are now defined in the most general
way.  Later, however, we will consider special forms of them that
allow us to simplify the relevant equations.  Note also that we take
the extinction $A$ at distance $D$ as a random variable.  Hence, we
have introduced above the probability distribution $p_A(A | D)$ rather
than a (deterministic) function $A(D)$.  This general approach can be
used to describe molecular clouds with ``patchy'' column densities,
which seems to be quite common \citep[see,
e.g.,][]{1999ApJ...512..250L, 1999A&A...345..965C}.  Finally, we use a
general form for the photometric error that includes the common case
where the error depends on the magnitude of the star.  Moreover, we
also consider through $p_\epsilon(\hat m | m)$ the case of
\textit{undetected\/} objects, i.e.\ the case where a star of true
magnitude $m$ close to the detection limit produces no detectable
flux; in this case we write $\hat m = \mathrm{null}$.  Hence, the
quantity
\begin{equation}
  \label{eq:2}
  c(m) = 1 - p_\epsilon(\mathrm{null} | m)
\end{equation}
represents the \textit{completeness\/} of our observations for stars
with true magnitude $m$.

We assume that both $p_A(A | D)$ and $p_\epsilon(\hat m | m)$ are
normalized to unity, i.e.
\begin{align}
  \label{eq:3}
  \int_0^\infty p_A(A | D) \, \diff A & {} = 1 &
  \forall D & \; , \\
  \label{eq:4}
  p_\epsilon(\mathrm{null} | m) + \int_{-\infty}^\infty
  p_\epsilon(\hat m | m) \, \diff \hat m & {} = 1 & \forall m& \; .
\end{align}
Moreover, we take $\rho(M, D)$ to be normalized so that the quantity
\begin{equation}
  \label{eq:5}
  \diff N = D^2 \rho(M, D) \, \diff D \, \diff \Omega \, \diff M 
\end{equation}
represents the expected number of stars located at a distance in the
range $[D, D + \diff D]$ in a patch of the sky of solid angle $\diff
\Omega$, and with absolute magnitudes between $M$ and $M + \diff M$.

Equation~\eqref{eq:1} can be used to obtain the probability
distribution of reddened magnitudes $p_m(m)$.  Using the previous
definitions we find
\begin{align}
  \label{eq:6}
  p_m(m) = \int_0^\infty & \diff D \, D^2 \int_0^\infty \diff A \,
  p_A(A | D) \notag\\
  & {} \times \rho(m - 5 \Log D + 5 - A, D) \; .
\end{align}
Finally, $p_m(m)$ can be converted into the distribution of observed
magnitudes $p_{\hat m}(\hat m)$ by convolving it with the measurement
error probability distribution $p_\epsilon(\hat m | m)$:
\begin{equation}
  \label{eq:7}
  p_{\hat m}(\hat m) = \int_\infty^\infty p_\epsilon(\hat m | m)
  p_m(m) \, \diff m \; .
\end{equation}
Because of the way $p_\epsilon(\hat m | m)$ is defined,
Eq.~\eqref{eq:7} includes also the case $\hat m = \mathrm{null}$.
Note also that $p_{\hat m}(\hat m)$ is normalized so that $p_{\hat
  m}(\hat m) \, \diff \hat m$ represents the expected angular density
of stars with observed magnitudes on the range $[\hat m, \hat m +
\diff \hat m]$.

Equations~\eqref{eq:6} and \eqref{eq:7} are basic equations of this
paper.  However, in the form they are written, they are by far too
general to be useful in practical cases.

\subsection{Simplifications}
\label{sec:simplifications}

In order to make use of the general relations written above we
introduce a number of useful simplifications.

First, we note that so far we have considered observations carried out
on a \textit{single\/} band.  In order to threat multi-band data, we
introduce a \textit{reddening law\/} as follows: for any band
$\lambda$, the extinction $A_\lambda$ is given by
\begin{equation}
  \label{eq:8}
  A_\lambda = k_\lambda A_V \; ,
\end{equation}
where $k_\lambda$ is a constant and $A_V$ is the extinction in the $V$
band.  In other words, Eq.~\eqref{eq:8} states that the ratio
$A_\lambda / A_{\lambda'}$ between the extinction in two bands
$\lambda$ and $\lambda'$ is a constant for a given cloud.  In reality,
this ratio is known to vary with the environment for short
wavelengths, but is almost universal in the NIR (and, according to a
recent work of \citealp{2004astro.ph..6403I}, for wavelengths in the
range $3$--$10 \ \mu\mbox{m}$ as well).  In the rest of this paper we
will follow the usual notation and express all extinctions in terms of
$A_V$.

A second important assumption that we will use regards the functional
form of $\rho(M, D)$.  First, we will assume that this function can be
separated in $M$ and $D$, i.e.\ can be written as the product of two
functions, one involving $M$ only, and one involving $D$ only:
$\rho(M, D) = p_M(M) \rho(D)$.  In practice, this assumption is
equivalent to saying that we observe a single ``population'' of stars
at all distances.  Since observations show that star number counts are
well approximated by an exponential law of the form $p_m(m) \propto
10^{\alpha m}$ (where the exponent $\alpha \simeq 0.34$ is
approximately independent of the band considered in the NIR), we also
assume that $p_M(M)$ follows a similar law.  We then write
\begin{equation}
  \label{eq:9}
  \rho(M, D) = \nu_\lambda 10^{\alpha M} \rho(D) \; ,
\end{equation}
where $\nu_\lambda$ is a normalization factor.  In the following we
will assume that $\rho(D)$, similarly to $\alpha$, is independent of
the band considered, while $\nu_\lambda$ is not.  Note also that,
because of the exponential form of this equation in $M$, a change of
the normalization factor $\nu_\lambda$ is in practice equivalent to a
change of the zero-point used for the magnitude $M$.

Using Eqs.~\eqref{eq:9} and \eqref{eq:8} we can slightly simplify
Eq.~\eqref{eq:6} by performing the integration over $D$:
\begin{align}
  \label{eq:10}
  p_m(m) = {} & \int_0^\infty \diff D \, D^{2 - 5 \alpha} \rho(D)
  \int_0^\infty \diff A_V \, p_{A_V}(A_V | D) \notag\\
  & {} \phantom{\int_0^\infty} \times \nu_\lambda 10^{5 \alpha}
  10^{\alpha m} 10^{-\alpha k_\lambda A_V} \notag\\
  {} \equiv {} & a \nu_\lambda 10^{\alpha m} \int_0^\infty \diff A_V
  \, p_{A_V}(A_V) 10^{-\alpha k_\lambda A_V} \; ,
\end{align}
where we have defined a distance-weighted probability distribution for
$A_V$ as
\begin{equation}
  \label{eq:11}
  p_{A_V}(A_V) = \frac{10^{5 \alpha}}{a} \int_0^\infty \diff D \,
  D^{2-5\alpha} \rho(D) p_{A_V}(A_V | D) \; ,
\end{equation}
and where $a$ is a numerical factor introduced in Eq.~\eqref{eq:10} to
ensure that $p_{A_V}(A_V)$ is normalized to unity:
\begin{equation}
  \label{eq:12}
  a = 10^{5\alpha} \int_0^\infty \diff D \, D^{2 - 5\alpha} \rho(D) \;
  .
\end{equation}
Note that $p_{A_V}(A_V)$ has a simple interpretation, being the
probability distribution for stars to have undergone a given
extinction regardless of their distance.  For example, if all stars
were subject to the same extinction $A_V$, we would have
$p_{A_V}(A_V') = \delta(A_V' - A_V)$.  Note also that, since $\alpha
\simeq 0.34$ can be taken to be approximately independent of the band
considered, $p_{A_V}(A_V)$ is also band-independent.

Equation~\eqref{eq:10} shows an interesting property of extinction
studies: the probability distribution $p_m(m)$ depends on $p_{A_V}(A_V
| D)$ only through $p_{A_V}(A_V)$.  This point is particularly
important since it shows an intrinsic limitation of extinction
measurements.  Indeed, all observables are derived from $p_{\hat
  m}(\hat m)$, the distribution of observed magnitudes, and this
function depends only on $p_m(m)$ (other than observational ``limiting
factors'' such as $p_\epsilon(\hat m | m)$ or $\hat c(\hat m)$).
Since, as noted above, $p_m(m)$ does not depend directly on
$p_{A_V}(A_V | D)$, any estimator based on observed magnitudes
only\footnote{Through this paper we assume that the distance of
  individual stars is not an observable; if, instead, the distance of
  each star can be estimated, then in principle one can also directly
  measure $p_{A_V}(A_V | D)$.} will not provide any information on
$p_{A_V}(A_V | D)$ directly but (in the best case scenario) only on
$p_{A_V}(A_V)$.  Since $p_{A_V}(A_V)$ is a sort of convolution of
$p_{A_V}(A_V | D)$, in general it is not possible to have a complete
knowledge on $p_{A_V}(A_V | D)$, not even if we know the star density
distribution $\rho(D)$; this limitation, among other things, prevents
us from gathering information on the three-dimensional structure of a
molecular cloud.

A case where, instead, the function $p_{A_V}(A_V | D)$ can be
recovered is a \textit{deterministic model\/} for the extinction,
i.e.\ a cloud complex whose extinction $A_V$ depends directly on the
distance $D$ (and, thus, is not a random variable).  In this case we
have
\begin{equation}
  \label{eq:13}
  p_{A_V}(A_V | D) = \delta \bigl( A_V - A_V(D) \bigr) \; .
\end{equation}
The extinction $A_V(D)$ has a simple relationship with the integrated
gas column density $\rho_\mathrm{gas}$:
\begin{equation}
  \label{eq:14}
  A_V(D) = \frac{1}{\beta} \int_0^D \diff D' \, \rho_\mathrm{gas}(D') \; ,
\end{equation}
where $\beta \simeq 2 \times 10^{21} \mbox{ cm}^{-2} \mbox{ mag}^{-1}$
is the ratio between gas density and $V$-band extinction
\citep{1955ApJ...121..559L, 1978ApJ...224..132B}.  Suppose now that we
carry out observations on the cloud and measure $p_{A_V}(A_V)$.  In
order to show that we can recover the structure of the cloud, let us
call $D(A_V)$ the inverse of the function $A_V(D)$ defined in
Eq.~\eqref{eq:14}, i.e.\ the distance at which the integrated
extinction is $A_V$.  Then we have
\begin{align}
  \label{eq:15}
  p_{A_V}(A_V) {} & = \frac{10^{5\alpha}}{a} \bigl[ D(A_V)
  \bigr]^{2-5\alpha} \rho\bigl( D(A_V) \bigr) D'(A_V) \notag\\
  {} & = \frac{10^{5\alpha}}{a} \frac{\rho\bigl( D(A_V) \bigr)}{3 -
    5\alpha} \frac{\diff}{\diff A_V} \Bigl[ \bigl[D(A_V)
  \bigr]^{3-5\alpha} \Bigr] \; .
\end{align}
In other words, we have obtained a differential equation in
$\bigl[D(A_V) \bigr]^{3-5\alpha}$.  If $\rho(D)$ is known, using the
boundary condition $D(0) = 0$ we can in principle solve this
differential equation and obtain $D(A_V)$ which, in turn, can be
inverted into $A_V(D)$ and $\rho_\mathrm{gas}(D)$.  It is superfluous
to stress that this process in reality is far from being trivial.

Finally, we consider a very simple deterministic model for
$p_{A_\lambda}(A_\lambda | D)$ that describes a thin cloud at a
distance $D_0$ with uniform column density $A_V$:
\begin{equation}
  \label{eq:16}
  p_{A_V}(A_V' | D) = 
  \begin{cases}
    \delta(A_V') & \text{if $D < D_0 \;, $} \\
    \delta(A_V' - A_V) & \text{otherwise$\; .$}
  \end{cases}
\end{equation}
The related distribution $p_{A_V}(A_V)$ takes then the simple form
\begin{equation}
  \label{eq:17}
  p_{A_V}(A_V') = f \delta(A_V') + (1 - f) \delta(A_V' - A_V) \; ,
\end{equation}
where $f$ is a real number in the range $[0, 1]$ given by
\begin{equation}
  \label{eq:18}
  f = \biggl[ \int_0^{D_0} \!\! \diff D \, D^{2 - 5\alpha} \rho(D)
  \biggr] \biggm/ \biggl[ \int_0^\infty \!\! \diff D \, D^{2 - 5\alpha}
  \rho(D) \biggr] \; .
\end{equation}
Hence, $f$ represents the fraction of foreground stars (i.e.\ stars at
distance $D < D_0$) present in any apparent magnitude bin in regions
with negligible extinction.  In the following, we will refer to the
simple case described in Eqs.~\eqref{eq:16} and \eqref{eq:17} as
``thin cloud approximation'' (the term is borrowed from gravitational
lensing theory).

\subsection{The completeness function}
\label{sec:compl-funct}

Above, in Eq.~\eqref{eq:2}, we introduced the completeness function
$c(m)$, which gives the probability to \textit{detect\/} a star of
magnitude $m$.  Typically, $c(m)$ is close to unity for bright stars,
and vanishes for very faint stars; note however that $c(m)$ might be
smaller than unity at low $m$ in crowded fields or close to bright
objects.  The completeness function $c(m)$ enters naturally in the
definition of the error probability distribution $p_\epsilon(\hat m |
m)$, which can be written as
\begin{equation}
  \label{eq:19}
  p_\epsilon(\hat m | m) = 
  \begin{cases}
    c(m) p_\epsilon(\hat m | m) & \text{if $\hat m \neq \mathrm{null}
      \; ,$} \\
    1 - c(m) & \text{if $\hat m = \mathrm{null} \; .$}
  \end{cases}
\end{equation}
This representation reflects the measurement process: for any star,
there is first a random process that ``decides'' whether the object is
detected or not (with probability fixed by $c(m)$); then, if
the star is detected, there is a second random process that generates
the observed magnitude $\hat m$ according to $p_\epsilon(\hat m | m)$.
Note also that this latter probability distribution is normalized
[cf.\ Eq.~\eqref{eq:4}].

In order to carry out some simplifications (see below
Sect.~\ref{sec:implementation}), we also consider a different
representation of the completeness function in which the order of the
two random processes described above is swapped: this leads to a
completeness function $\hat c(\hat m)$ defined in terms of the
\textit{observed\/} magnitude.  In other words, we first generate for
\textit{every\/} star the ``observed'' magnitude $\hat m$ according to
a probability distribution $\hat p_\epsilon(\hat m | m)$, and then
decide whether the star is really observed using a second random
process controlled by the completeness function $\hat c(\hat m)$.
This implies, among other things, a modification of Eq.~\eqref{eq:7},
that now becomes
\begin{equation}
  \label{eq:20}
  p_{\hat m}(\hat m) = \hat c(\hat m) \int_{-\infty}^\infty
  \hat p_\epsilon(\hat m | m) p_m(m) \, \diff m \; .
\end{equation}
This equation is valid for detected stars, i.e.\ if $\hat m \neq
\mathrm{null}$.  Since $\hat c(\hat m)$ represents the probability
that a star with \textit{measured\/} magnitude $\hat m$ be detected,
we evaluate the probability that the star is \textit{not\/} detected
as
\begin{equation}
  \label{eq:21}
  p_{\hat m}(\mathrm{null}) = \int_{-\infty}^{\infty} \! \diff \hat m
  \, \bigl[ 1 - \hat c(\hat m) \bigr]  \int_{-\infty}^{\infty} \!
  \diff m \, \hat p_\epsilon(\hat m | m) p_m(m) \; .
\end{equation}
An important point to observe here is the fact that we have modified
Eq.~\eqref{eq:7} into Eqs.~\eqref{eq:20} and \eqref{eq:21} by defining
\textit{two\/} functions, the new error distribution $\hat
p_\epsilon(\hat m | m)$ and the new completeness function $\hat c(\hat
m)$, which replace, respectively, $p_\epsilon(\hat m | m)$ and $c(m)$.
Indeed, both these couples of functions are intimately related, and
this was implicitly taken into account above by using a single
distribution $p_\epsilon(\hat m | m)$ to describe both the photometric
errors and the completeness of the observations.  Interestingly, it is
possible to find a relationship between the couple $\hat
p_\epsilon(\hat m | m)$, $\hat c(\hat m)$ and the couple
$p_\epsilon(\hat m | m)$, $c(m)$ by requiring that, for any reddened
magnitude probability distribution $p_m(m)$, the observed magnitude
probability distribution $p_{\hat m}(\hat m)$ evaluated using
Eq.~\eqref{eq:7}, or using Eqs.~\eqref{eq:20} and \eqref{eq:21},
agrees.  We find then
\begin{align}
  \label{eq:22}
  c(m) p_\epsilon(\hat m | m) & {} = \hat c(\hat m) \hat
  p_\epsilon(\hat m | m) \; , \\
  \label{eq:23}
  1 - c(m) & {} = \int_{-\infty}^\infty  \hat p_\epsilon(\hat m |
  m) \bigl[ 1 -  \hat c(\hat m) \bigr] \, \diff \hat m \; .
\end{align}
Since $p_\epsilon(\hat m | m)$ is normalized to unity, if we
integrate Eq.~\eqref{eq:22} over $\hat m$ and substitute the result
into Eq.~\eqref{eq:23}, we find
\begin{equation}
  \label{eq:24}
  \int_{-\infty}^\infty \hat p_\epsilon(\hat m | m) \, \diff \hat
  m = 1 \; ,
\end{equation}
i.e.\ $\hat p_\epsilon$ is also normalized to unity.  In summary
we have:
\begin{align}
  \label{eq:25}
  c(m) & {} = \int_{-\infty}^\infty \hat c(\hat m) \hat
  p_\epsilon(\hat m | m) \, \diff \hat m \; , \\
  \label{eq:26}
  p_\epsilon(\hat m | m) & {} = \frac{\hat c(\hat m) \hat
    p_\epsilon(\hat m | m)}{\int_{-\infty}^\infty \hat c(\hat m) \hat
    p_\epsilon(\hat m | m) \, \diff \hat m} \; .
\end{align}
Since Eq.~\eqref{eq:25} involves a convolution, it is in general not
possible to invert it and express $\hat c(\hat m)$ in terms of $c(m)$.
Note that Eq.~\eqref{eq:26} is essentially Bayes' theorem applied to
$p_\epsilon(\hat m | m)$.

Although Eqs.~\eqref{eq:25} and \eqref{eq:26} show that the
description of the completeness $c(m)$ in terms of the true reddened
magnitude $m$ is more general than the description $\hat c(\hat m)$ in
terms of the observed magnitude $\hat m$, we argue that the latter is
more practical to use:
\begin{itemize}
\item Operationally, the completeness function is usually evaluated
  from the \textit{observed magnitudes\/} by comparing the expected
  number of stars in a given magnitude bin with the number of stars
  detected in the same bin.
\item Often, it can be sensible to use some \textit{a posteriori\/}
  cuts in the star catalog.  For example, if we know that observations
  of faint stars are particularly unreliable, we can just discard all
  objects with observed magnitude $\hat m$ larger than a given
  threshold $m^\mathrm{lim}$.  This cut could be easily described in
  terms of $\hat c(\hat m)$:
  \begin{equation}
    \label{eq:27}
    \hat c(\hat m) = H(m^\mathrm{lim} - \hat m) = 
    \begin{cases}
      1 & \text{if $\hat m \le m^\mathrm{lim} \; ,$} \\
      0 & \text{otherwise$\; .$}
    \end{cases}
  \end{equation}
\item The use of $\hat c(\hat m)$ allows us to evaluate analytically
  $p_{\hat m}(\hat m)$ for some special simple probability
  distributions.  This point is particularly valuable for the
  practical applications that we will describe below in
  Sect.~\ref{sec:likelihood-approach}.
\end{itemize}

\subsection{Multi-band probability distributions}
\label{sec:multi-band-prob}

So far we have focused our analysis on single-band measurements.  In
order to extend the discussion to observations carried out in
different bands, we need to generalize the relevant probability
distributions.

We denote $\vec M = \{ M_1, M_2, \dots, M_\Lambda \}$ the magnitudes
of a star in $\Lambda$ different bands; in general, we use bold
symbols such as $\vec k$ to indicate quantities that have different
values in the various bands.  We will threat these as vector
quantities; in this way, e.g., the generalization of Eq.~\eqref{eq:1}
can written as
\begin{equation}
  \label{eq:28}
  \hat{\vec m}_n = \vec M_n + 5 \Log D_n - 5 + \vec k A_V + \vec
  \epsilon_n = \vec m_n + \vec \epsilon_n \; .
\end{equation}

Since we are now working with observations in different bands, we need
also to generalize $\rho(M, D)$.  We define thus $\rho(\vec M, D)$,
the probability to have a star with absolute magnitudes $\vec M$ at
distance $D$ from us.  Using this distribution we can rewrite
Eq.~\eqref{eq:6} as
\begin{align}
  \label{eq:29}
  p_m(\vec m) = \int_0^\infty & \diff D \, D^2 \int_0^\infty \diff A_V
  \, p_{A_V}(A_V | D) \notag\\
  & {} \times \rho(\vec m - 5 \Log D + 5 - \vec k A_V, D) \; .
\end{align}

The generalization of Eq.~\eqref{eq:7} is also simple.  Since
photometric measurements in different bands are independent, we have
to perform $\Lambda$ different convolutions with the measurement error
probability distributions $\bigl\{ p_{\epsilon_\lambda}(\hat m_\lambda
| m_\lambda) \bigr\}$ corresponding to the various bands:
\begin{equation}
  \label{eq:30}
  p_{\hat m}(\hat{\vec m}) = \int \diff^\Lambda \! m \, p_m(\vec m)
  \prod_{\lambda=1}^\Lambda p_{\epsilon_\lambda}(\hat m_\lambda |
  m_\lambda) \; .
\end{equation}
Note that we consider a star detected if it is detected in \textit{at
  least\/} one band.

The integral of $p_{\hat m}(\hat{\vec m})$ over all admissible values
of $\hat{\vec m}$ gives the expected local density of stars $\sigma$:
\begin{equation}
  \label{eq:31}
  \sigma = \int_{( \R \cup \{ \mathrm{null} \} )^\Lambda \setminus \{
  \mathbf{null} \}} p_{\hat m}(\hat{\vec m}) \, \diff^\Lambda \! \hat
  m \; . 
\end{equation}
Since $\sigma$ gives the normalization of $p_{\hat m}(\hat{\vec m})$,
the conditional probability that a star with magnitudes $\hat{\vec m}$
be observed given the fact that the star is detected is $p_{\hat
  m}(\hat{\vec m}) / \sigma$.

When using the alternate completeness functions $\bigl\{ \hat
c_\lambda(\hat m_\lambda) \bigr\}$ expressed in terms of the observed
magnitudes $\{ \hat m_\lambda \}$, Eq.~\eqref{eq:30} can be rewritten
as
\begin{align}
  \label{eq:32}
  p_{\hat m}(\hat{\vec m}) = {} & \int \diff^\Lambda \! \hat m' \,
  \prod_{\lambda=1}^\Lambda \Bigl[ \delta(\hat m_\lambda - \hat
  m'_\lambda) \hat c_\lambda(\hat m'_\lambda) \notag\\
  & \phantom{\int \diff^\Lambda \! \hat m' \,
    \prod_{\lambda=1}^\Lambda \Bigl[} + \delta(\hat m_\lambda -
  \mathrm{null}) \bigl( 1 -
  \hat c_\lambda(\hat m'_\lambda) \bigr) \Bigr] \notag\\
  & {} \times \int \diff^\Lambda \! m \, p_m(\vec m)
  \prod_{\lambda=1}^\Lambda p_{\epsilon_\lambda}(\hat m_\lambda' |
  m_\lambda) \; .
\end{align}
The combination of delta distributions inside the brackets in this
equation ensures that $\hat m'_\lambda = \hat m_\lambda$ if $\hat
m_\lambda \neq \mathrm{null}$, and that we integrate over $\bigl( 1 -
\hat c_\lambda(\hat m_b') \bigr)$ (the probability of \textit{not\/}
detecting the star) if $\hat m_\lambda = \mathrm{null}$; the last
integration is the usual convolution with the measurement errors.
Similarly, Eq.~\eqref{eq:31} can also be written as
\begin{align}
  \label{eq:33}
  \sigma = {} & \int \diff^\Lambda \! \hat m \biggl[ 1 -
  \prod_{\lambda=1}^\Lambda \Bigl( 1 - \hat c_\lambda(\hat m_\lambda)
  \Bigr) \biggr] \notag\\ 
  & {} \times \int \diff^\Lambda \! m \, p_m(\vec m)
  \prod_{\lambda=1}^\Lambda p_{\epsilon_\lambda}(\hat m_\lambda |
  m_\lambda) \; .
\end{align}

\subsection{Further simplifications}
\label{sec:furth-simpl}

As for the single-band case, we consider a number of realistic and useful
simplifications that will allow us to take advantage of the formalism
introduced above in practical cases.

First, we still take $\rho(\vec M, D)$ to be separable into $\rho(\vec
M, D) = p_M(\vec M) \rho(D)$.  Furthermore, we adopt for $p_M(\vec M)$
a simple functional form, which is sufficiently accurate for our
purposes.  Since we know that $p_m(m) \propto 10^{\alpha m}$ with
$\alpha$ approximately independent of the band, and since stars appear
to have a limited scatter in their NIR colors, we write
\begin{multline}
  \label{eq:34}
  p_M(\vec M) = \nu 10^{\alpha M_1} \\
  {} \times \exp \left[ -\frac{(M_a - M_1 - \chi_a) C^{-1}_{ab} (M_b -
      M_1 - \chi_b)}{2} \right] \; ,
\end{multline}
where we have used Einstein's convection on repeated indexes.  In
other words, we suppose that $M_1$ is exponentially distributed, and
that star colors are distributed as Gaussian random variables with
averages $\langle M_a - M_1 \rangle = \chi_a$; $C$ represents the
covariance of these colors.  This distribution can be rewritten in a
more convenient way as (see also below Sect.~\ref{sec:implementation})
\begin{equation}
  \label{eq:35}
  p_M(\vec M) = \exp \left[ - \frac{\vec M^\mathrm{T} P \vec M + 2
  \vec Q^\mathrm{T} \vec M + R}{2} \right] \; .
\end{equation}
The form \eqref{eq:34} for $p_M(\vec M)$ is particularly convenient
for several reasons.  First, we can greatly simplify Eq.~\eqref{eq:29}
and obtain a result similar to Eq.~\eqref{eq:10}:
\begin{equation}
  \label{eq:36}
  p_m(\vec m) = a \int_0^\infty \diff A_V \, p_{A_V}(A_V) p_M(\vec m -
  \vec k A_V) \; ,
\end{equation}
where $p_{A_V}(A_V)$ and $a$ are still given by Eqs.~\eqref{eq:11} and
\eqref{eq:12}.  Again, we observe that the fact that $p_m(\vec m)$
depends only on the distance-weighted $p_{A_V}(A_V)$, implies that
$p_{A_V}(A_V | D)$ is not an observable.  Moreover, the forms
\eqref{eq:34} and \eqref{eq:35} are unmodified by a reddening $\vec M
\mapsto \vec m = \vec M + 5 \Log D - 5 + \vec k A_V$ and by a
convolution with Gaussian photometric errors (see below
Sect.~\ref{sec:implementation}).

\section{Extinction measurements}
\label{sec:extinct-meas}

We will describe in this section the star counts and color excess
methods in order to better describe the advantages and limitations of
them.

\subsection{Star counts}
\label{sec:star-counts}

In the 18th century the English astronomer William Herschel noted that
some regions presented few stars and, following Newton's idea of a
perfectly transparent space, interpreted these ``holes in the sky'' as
a real lack of stars.  This misconception survived the discovery of
individual dark clouds \citep{1919ApJ....49....1B} and had serious
consequences on Shapley's calibration of the distance scale for
Cepheids.  The ``discovery'' of dust in our Galaxy took place only in
\citeyear{1930LicOB..14..154T} when \citeauthor{1930LicOB..14..154T}
showed its importance in dimming the light coming from distant open
clusters.  Finally, in recent decades dust has no longer been seen
only as annoying ``fog'' obscuring the light of background sources,
and has been shown to have a tremendous impact on the evolution of
galaxies and on the formation of stars and stellar systems (see
\citealp{2003ssac.proc...37L} for a detailed historical review).

It has long been recognized \citep{1923AN....219..109W} that
measurements of the local density of stars in different regions of the
sky can be used to map the extinction.  The original technique
consisted in comparing the number of stars in magnitude bins in
regions subject to extinction with the number of stars in regions
where the extinction is (supposedly) negligible.  This technique was
then improved by \citet{1956AJ.....61..309B} which suggested to use
counts up to a limiting magnitude to reduce the error.  In the past,
the star count technique was mainly applied to optical data (typically
visually inspected Smith plates; e.g.  \citealp{1978AJ.....83..363D};
\citealp{1986A&A...160..157M}; \citealp{1996A&AS..116...21A}).  In the
last decade, however, near-infrared digital data have been available,
and the star count technique has been finally applied to NIR
observations \citep[e.g.][]{1997A&A...324L...5C}.  In this respect, a
key role has been played by large NIR surveys such as the Two Micron
All Sky Survey (2MASS; \citealp{1994ExA.....3...65K}) and the Deep NIR
southern Sky Survey (DENIS; \citealp{1997Msngr..87...27E}).

The star count method is easily described using our notation.  We
first note that, as for multi-band probability distributions, the
integral of $p_{\hat m}(\hat m)$ over $\hat m$ gives the local density
of stars $\sigma$ [cf.\ Eq.~\eqref{eq:31}]:
\begin{align}
  \label{eq:37}
  \sigma \equiv \langle \sigma \rangle = {} & \int p_{\hat m}(\hat m)
  \, \diff \hat m \notag\\ 
  {} = {} & \int \diff \hat m \, \hat c(\hat m) \int \diff m \, p_m(m)
  p_\epsilon(\hat m | m) \; .
\end{align}
Inserting here Eq.~\eqref{eq:10}, we immediately obtain
\begin{equation}
  \label{eq:38}
  \sigma = \sigma^{(0)} \int \diff A_V \, p_{A_V}(A_V) 10^{-\alpha
  k_\lambda A_V} \; ,
\end{equation}
where $\sigma^{(0)}$ is the average density expected in absence of
extinction:
\begin{equation}
  \label{eq:39}
  \sigma^{(0)} = a \nu_\lambda \int \diff \hat m \, \hat c(\hat m)
  \int \diff m \, 10^{\alpha m} p_\epsilon(\hat m | m) \; .
\end{equation}
Equation~\eqref{eq:38} shows that the ratio $ \sigma / \sigma^{(0)}$
between the average densities expected in presence and in absence of
extinction is simply related to $p_{A_V}(A_V)$.

Clearly, a single measurement of $ \sigma / \sigma^{(0)}$ cannot be
used to derive the distribution $p_{A_V}(A_V)$.  However, in the
simplest case where all stars are background to a thin cloud, so that
$p_{A_V}(A'_V) = \delta(A_V - A'_V)$, we find the classical relation
\begin{equation}
  \label{eq:40}
  \frac{\sigma}{\sigma^{(0)}} = 10^{- \alpha k_\lambda A_V} \; ,
\end{equation}
which can be immediately inverted to obtain $A_V$
\citep{1956AJ.....61..309B}.  Operationally, both densities $\sigma$
and $\sigma^{(0)}$ are measured by dividing the number of stars
observed against the cloud and in a control field by the angular size
of the regions considered.  Hence, the measurement errors on these
quantities are due to the randomness on the local number of stars.  If
we ignore the correlation on the star positions, and thus assume that
these are a homogeneous Poisson process \citep[see, e.g.,][]{Cressie},
then the local number of stars follows a simple Poisson distribution.

\begin{table}[t]
  \centering
  \begin{tabular}{lcc}
    $\lambda $ & $k_\lambda$ & $k_\lambda$ \\
    & (Rieke \& Lebofsky 1985) & (van de Hulst 1946) \\
    \hline
    $J$ & 0.282 & 0.246 \\
    $H$ & 0.175 & 0.155 \\
    $K$ & 0.112 & 0.089 \\
    $L$ & 0.058 & 0.045 \\
    $M$ & 0.023 & 0.033 \\
    $N$ & 0.052 & 0.013
  \end{tabular}
  \caption{The extinction law in different infrared bands (taken from
  \citealp{1985ApJ...288..618R} and \citealp{1946RAOU...11....2V}).}
  \label{tab:1}
\end{table}

Although the estimator \eqref{eq:40} can be applied to thin clouds
only, in principle more general situations can also be handled by
taking advantage of the particular form of Eq.~\eqref{eq:38}.  We
first observe that the integral appearing in the r.h.s.\ of this
equation is reminiscent of a Laplace's transform.  Given a real
function $f(x)$ defined for non-negative values $x$, its Laplace's
transform is defined as
\begin{equation}
  \label{eq:41}
  \mathcal{L}[f](k) = \int_0^\infty f(x) \exp( - k x) \; .
\end{equation}
An important property of Laplace's transform is that it can be
inverted: in other words, it is possible to obtain $f(x)$ provided one
knows $\mathcal{L}[f](k)$ for any non-negative $k$.  Using the
notation just introduced, we can rewrite Eq.~\eqref{eq:38} as
\begin{equation}
  \label{eq:42}
   \sigma = \sigma^{(0)} \mathcal{L}[p_{A_V}](\alpha k_\lambda \ln 10)
   \; .
\end{equation}
Hence, if we were able to measure the ratio $\sigma / \sigma^{(0)}$
for all positive values of the product $\alpha k_\lambda$, we could in
principle derive $p_{A_V}$.  As indicated above, the constant $\alpha
\simeq 0.34$ is approximately independent of the band considered;
however, $k_\lambda$ strongly depends on the band used to carry out
the observations (see Table~\ref{tab:1}).  Hence, the product $\alpha
k_\lambda$ that appears in Eq.~\eqref{eq:42} can be varied among a
\textit{limited\/} set of values.  In conclusion, although complete
knowledge of $p_{A_V}$ is clearly impossible, we can still use
multiband observations to investigate $p_{A_V}$ in more complex
situations than the one considered in Eq.~\eqref{eq:40}.

Let us consider, for example, the case of the thin cloud approximation
described in Eq.~\eqref{eq:17}.  Since, in general, the fraction $f$
of foreground stars is not known, we want to estimate both $A_V$ and
$f$ in a given patch of the sky. [Although $f$ can be taken to be
constant in many cases, changes on this quantity have to be expected
for different regions of large cloud complexes because of the geometry
of the cloud and of changes of star densities due to the Galactic
structure.]\@ For this, we insert Eq.~\eqref{eq:17} in
Eq.~\eqref{eq:38}, thus obtaining a simple generalization of
Eq.~\eqref{eq:40}:
\begin{equation}
  \label{eq:43}
  \frac{\sigma}{\sigma^{(0)}} = f + (1 - f) 10^{- \alpha k_\lambda
  A_V} \; . 
\end{equation}
Hence, we can in principle use two different measurements of
$\bar\sigma$ in two different bands to deduce both $f$ and $A_V$ on
the region considered.  In practice, it is preferable to follow the
approach described below in Sect.~\ref{sec:likelihood-approach}.

\begin{figure}[t]
  \centering
  \includegraphics[width=\hsize,keepaspectratio]{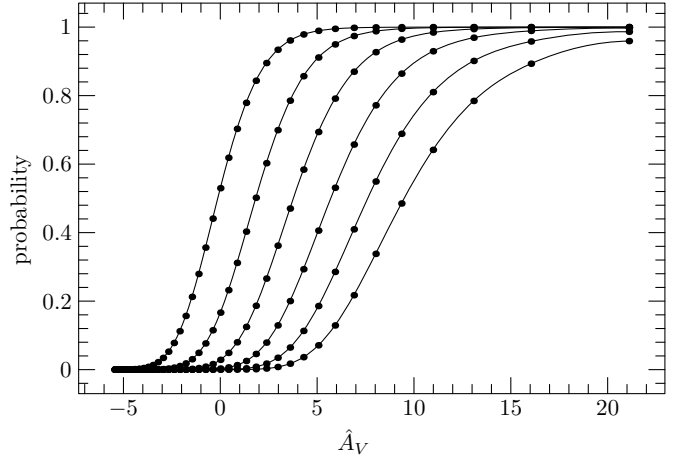}
  \caption{The cumulative probability distribution for $\hat A_V$.
    The various curves are relative to a true extinction $A_V =
    [0,2,4,6,8,10]$ (from the left to the right); the average values
    measured are instead $\langle \hat A_V \rangle = [0.24, 2.33,
    4.47, 6.64, 8.79, 10.76]$ (the scatters around these values ranges
    from $1.9$ to $4.3$).  In all cases we assumed $\alpha = 0.34$,
    $k_\lambda = 0.175$, $f = 0.1$, and $\mathcal{A} \sigma^{(0)} =
    20$.  Along the curves the dots mark the actual possible values of
    $A_V$ that can be measured for different values of the observed
    number of stars $N$.}
  \label{fig:1}
\end{figure}

It is interesting to investigate in more detail the estimator derived
from Eq.~\eqref{eq:43}, i.e.
\begin{equation}
  \label{eq:44}
  \hat A_V = - \frac{1}{\alpha k_\lambda} \Log \left[ \frac{N -
  \mathcal{A} \sigma^{(0)} f}{\mathcal{A} \sigma^{(0)} (1 - f) }
  \right] \; , 
\end{equation}
where $N$ is the number of stars found in the region investigated, and
$\mathcal{A}$ is its area.  Assuming that $\sigma^{(0)}$ is measured
without significant errors (this is possible by using a large control
field), we can deduce the expected error on $\hat A_V$ by noting that
$N$ follows a Poisson distribution with average [cf.
Eq.~\eqref{eq:43}]
\begin{equation}
  \label{eq:45}
  \langle N \rangle = \mathcal{A} \sigma^{(0)} \bigl[ f + (1-f)
  10^{-\alpha k_\lambda A_V} \bigr] \; .
\end{equation}
Using a first order approximation (valid for small relative errors) we
find 
\begin{equation}
  \label{eq:46}
  \mathrm{Var}\bigl(\hat A_V \bigr) = \left( \frac{1}{\alpha
  k_\lambda \ln 10} \cdot \frac{1}{\langle N \rangle - \mathcal{A} f
  \sigma^{(0)}} \right)^2 \langle N \rangle \; .
\end{equation}
In case of vanishing $f$ (i.e., if all stars are background to the
cloud), this expression takes a simpler form
\begin{equation}
  \label{eq:47}
  \mathrm{Err} \bigl( \hat A_V \bigr) = \sqrt{\mathrm{Var}\bigl(\hat
  A_V \bigr)} = \frac{1}{\alpha k_\lambda \ln 10}
  \frac{1}{\sqrt{\langle N \rangle}} = \frac{11.4}{\sqrt{\langle N
  \rangle}}\; , 
\end{equation}
where the last expression is valid for the $K$ band (cf.\ 
Tab.~\ref{tab:1}).

\begin{figure}[t]
  \centering
  \includegraphics[width=\hsize,keepaspectratio]{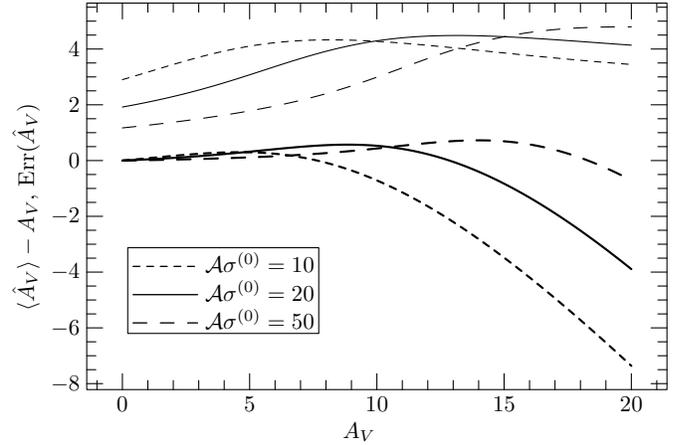}
  \caption{The bias (thick lines) and error (thin lines) on $\hat A_V$
    evaluated using Eq.~\eqref{eq:44} for different values of
    $\mathcal{A} \sigma^{(0)}$.  For this figure we used the same
    parameters as in Fig.\ref{fig:1}, i.e.\ $\alpha = 0.34$,
    $k_\lambda = 0.175$, and $f = 0.1$.}
  \label{fig:2}
\end{figure}

Further statistical properties of the estimator \eqref{eq:44} can be
better evaluated using numerical methods.  Figure~\ref{fig:1} shows,
as an example, the expected cumulative probability distributions for
$\hat A_V$ for some typical cases.  As expected, the cumulative
distributions reach $0.5$ around the true value for $A_V$, but the
expected measurement errors can be very large, especially at high
column densities.  The numerical analysis of Eq.~\eqref{eq:44} also
shows that the estimator is significantly biased.  In general, in the
limit $\sigma \mathcal{A} \gg 1$, the estimator is biased toward
large values of $\hat A_V$, i.e. $\bigl\langle \hat A_V \bigr\rangle >
A_V$; the opposite happens for $\sigma \mathcal{A} \sim 1$ (see
Fig.~\ref{fig:2}).  This change in behavior can be understood by
observing that, if $A_V$ is large, then $\mathcal{A} \sigma$ is
small and thus we simply do not have enough background stars to probe
high column densities.

Interestingly, Eq.~\eqref{eq:46} is a reasonably good approximation of
the true variance of $\hat A_V$.  Typically, this approximation
slightly underestimates the true variance of $\hat A_V$, but again the
opposite happens for low values of $\sigma \mathcal{A}$.

\subsection{Near-infrared color excess (N{\small ICE} and 
  N{\small ICER})} 
\label{sec:near-infrared-color}

The dust present in molecular clouds produces different amounts of
obscuration in different color bands (see Table~\ref{tab:1}), so that
background stars appear reddened.  Hence, the color excess in the red
part of the spectrum of background stars can be used to measure the
extinction along the line of sight.

Historically, the color excess technique has been impeded by the
limited instrumental sensitivity to small regions
\citep{1980ApJ...242..132J, 1982ApJ...262..590F, 1984ApJ...282..675J}.
However, in the early 1990s, with the advent of near-infrared arrays,
it became possible to accurately measure the NIR magnitudes of many
stars from single pointing observations.  This new technology was
first exploited by \citet{1994ApJ...429..694L}, and since then has
been successfully applied to many molecular clouds
\citep[e.g.][]{1997AJ....113.1788H, 1999ApJ...512..250L,
  2001Natur.409..159A}.

Lada's technique (called ``near-infrared color excess'' or
\textsc{Nice}) is based on measurements of a NIR color (e.g., $H - K$)
of many stars.  Since stars have relatively well defined colors in the
infrared, a significant intervening extinction can be detected as a
reddening.  Note that a key point of the \textsc{Nice} method (and of
similar methods based on the reddening of stars; see, e.g.,
\citealp{1999A&A...349L..69S}) is the assumption that all stars belong
to a homogeneous population.

The \textsc{Nice} method can be quantitatively described using the
notation introduced so far.  In particular, from Eq.~\eqref{eq:29} we
have 
\begin{equation}
  \label{eq:48}
  p_m(\vec m) = p^{(0)}_m(\vec m - \vec k A_V) \; ,
\end{equation}
where we have denoted $p^{(0)}_m(\vec m)$ the l.h.s.\ of
Eq.~\eqref{eq:29} for $A_V = 0$.  Naively, Eq.~\eqref{eq:48} can be
used to obtain a simple estimate of the intervening infrared
extinction $A_V$.  For example, we find
\begin{equation}
  \label{eq:49}
  \langle \vec m \rangle = \langle \vec m \rangle^{(0)} + \vec k A_V
  \; , 
\end{equation}
where again we used the superscript $(0)$ to denote quantities
measured in a control region where $A_V = 0$.  Equation~\eqref{eq:49}
is better rewritten in terms of star colors.  Calling $\chi_\lambda =
m_\lambda - m_1$ and $\kappa_\lambda = k_\lambda - k_1$, we have
\begin{equation}
  \label{eq:50}
  \langle \vec \chi \rangle = \langle \vec \chi \rangle^{(0)} + \vec
  \kappa A_V \; . 
\end{equation}
This equation, applied to a single color, is essentially the
\textsc{Nice} technique.  More precisely, this technique uses the
simple average of a set of $N$ angularly close stars to evaluate the
column density:
\begin{equation}
  \label{eq:51}
  \hat A_V = \frac{1}{N} \sum_{n=1}^N \frac{\hat \chi_n -
  \bar\chi^{(0)}}{\kappa} \; ,
\end{equation}
where $\bar \chi^{(0)}$ is an estimate of $\langle \chi \rangle^{(0)}$
(obtained, e.g., by measuring the star colors on a control field where
presumably $A_V \simeq 0$).  As an example, consider the
\textsc{Nice\/} method applied to the $\chi = H - K$ color.  Since
stars have a typical scatter of $0.09 \mbox{ mag}$ in this color, we
expect an error on $\hat A_V$ of $\mathrm{Err}\bigl( \hat A_V \bigr)
\simeq 1.4 / \sqrt{N}$.  Hence, even in the presence of significant
photometric errors, the \textsc{Nice} method gives significantly more
accurate results than the star count method [cf.\ Eq.~\eqref{eq:47}].

As shown by \citet{2001A&A...377.1023L}, one can indeed take full
advantage of observations carried out in different bands to obtain
more accurate column density measurements.  The improved technique,
called \textsc{Nicer} (\textsc{Nice} Revised) optimally balances the
information from different bands and different stars.  As a by-product
of the analysis, \textsc{Nicer} also allows us to evaluate the
expected error on the column density map, which is useful to estimate
the significance on the detection of substructures and cores.  The
\textsc{Nicer} technique can be described using the following simple
argument.  Equation~\eqref{eq:50} written above can be taken to be a
system of $(\Lambda - 1)$ equations to be approximately solved for
$A_V$, the approximation being made necessary because we can only
measure $\langle \vec \chi \rangle$ and $\langle \vec \chi
\rangle^{(0)}$ with limited accuracy.  The ``best'' solution for $A_V$
can been obtained by minimizing the chi-square quantity
\begin{equation}
  \label{eq:52}
  \chi^2 = \sum_{n=1}^N \bigl[ \hat{\vec \chi}_n - \bar {\vec
  \chi}^{(0)} - \vec \kappa A_V \bigr]^\mathrm{T} (C + E)^{-1} \bigl[
  \hat {\vec \chi}_n - \bar{\vec \chi}^{(0)} - \vec \kappa A_V \bigr]
  \; ,
\end{equation}
Consistently with the notation used above in Eq.~\eqref{eq:34}, we
called $C$ the covariance matrix of the star colors; moreover, the
symbol $E$ was used to denote the covariance matrix of measurement
errors [the two covariance matrices have to be added up in
Eq.~\eqref{eq:52} in order to properly estimate the expected scatter
on star colors].  The best estimate of $A_V$, obtained by minimizing
the $\chi^2$ of Eq.~\eqref{eq:52}, is precisely the \textsc{Nicer}
estimator.

Both the \textsc{Nice} and \textsc{Nicer} estimators \textit{appear\/}
to be unbiased provided that: (i) there are no foreground stars and
(ii) the measured colors $\hat \chi_n$ are unbiased estimates of the
true colors $\vec \chi_n$.

\begin{figure}[t]
  \centering
  \includegraphics[width=\hsize,keepaspectratio]{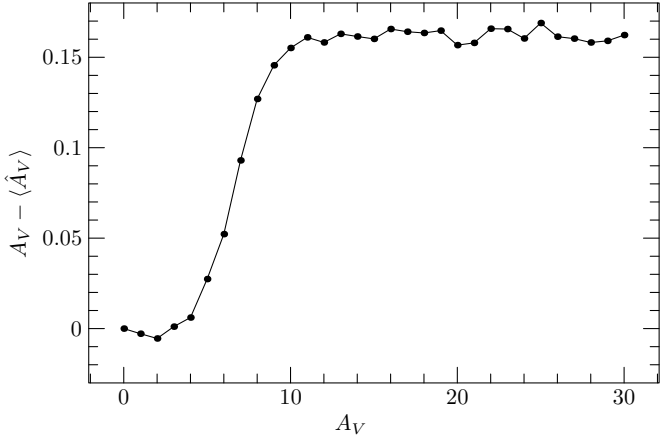}
  \caption{The bias for the \textsc{Nice} method due to the different
    completeness at low and high column densities.  This plot is
    evaluated by generating $100\,000$ stars for various values of
    $A_V$ (marked as dots).  Note that, as described in the text, for
    $A_V \simeq 6.7$ we observe a rapid increase in the bias due to
    the change between a selection in the $K$ band and a selection in
    the $H$ band (see Fig.~\ref{fig:4}).  The small scale oscillations
    on the plot are due to numeric effects.}
  \label{fig:3}  
\end{figure}

\begin{figure}[t]
  \centering
  \includegraphics[width=\hsize,keepaspectratio]{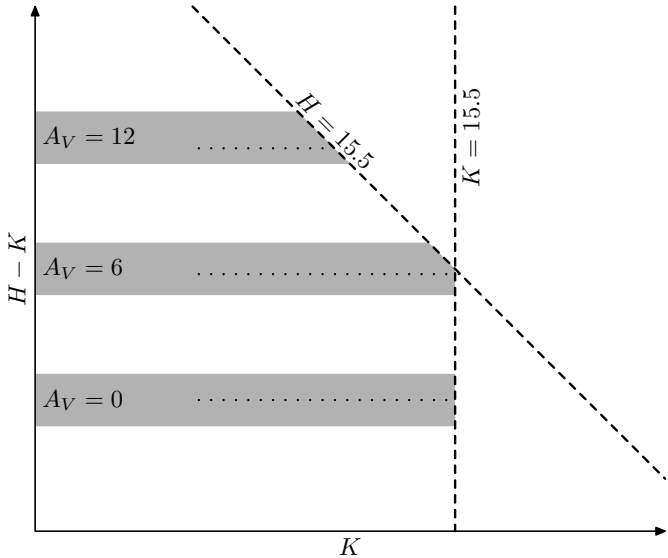}
  \caption{A graphic explanation of the bias plotted in
    Fig.~\ref{fig:3}.  At low column densities, we select stars mainly
    on their $K$ magnitude, while at high column densities we mainly
    select using their $H$ magnitude.  As a result, the average
    intrinsic color (dotted line) of the observed stars (gray bands)
    changes toward lower $H - K$ values.}
  \label{fig:4}  
\end{figure}

In reality, even if the two conditions considered above are satisfied,
both color excess methods can still be biased because of selection
effects introduced by the completeness function.  To understand this
point let us make a simple example.  Suppose that we carry out our
observations in two bands, $\lambda_1$ and $\lambda_2$, and that both
completeness functions $c_\lambda(m_\lambda)$ are not vanishing only
on a narrow magnitude range.  In this case we would always have $\hat
\chi \simeq \hat \chi^{(0)}$ (because a star is observed in both bands
only if $m_1$ and $m_2$ are inside the narrow detection window), and
thus we would always measure $\hat A_V \simeq 0$, independently of the
real column density.  Although unrealistic, this example shows that we
must expect a bias for the \textsc{Nice} method; the argument for the
\textsc{Nicer} method is similar and leads to the same conclusion.

The bias of these two methods depends on the details of the
probability distribution $p_M(\vec M)$ and of the completeness
functions $c_\lambda(m_\lambda)$.  However, as an example, we
evaluated the expected bias for various values of the column density
and for the typical probability distributions that we expect for the
2MASS catalog.  As shown in Fig.~\ref{fig:3}, the bias in the case
considered appears to be limited, below $0.2$ in $A_V$, and has a
characteristic shape: it vanishes for $A_V = 0$, increases quickly for
$A_V \sim 7$, and finally saturates for $A_V > 11$.  This behavior has
a simple explanation:
\begin{itemize}
\item Since the colors of unreddened stars are evaluated using a
  control field (where supposedly $A_V = 0$), the bias has to vanish
  for $A_V = 0$.
\item The general trend of bias on $\hat A_V$ can be understood with
  the help of Fig.~\ref{fig:4}.  At low $A_V$, stars that are observed
  in the $K$ band are almost certainly also observed on the $H$ band,
  because of the values of the limiting magnitudes (approximately
  $14.9$ in $H$ and $14.3$ in $K$) and of the average star colors
  ($\langle H - K \rangle = 0.18$).  When, instead, the reddening is
  large, we have the opposite situation (because $K$ is less affected
  by reddening than $H$).  This different selection at different
  column densities is the source of the observed bias.
\item Finally, at large column densities an asymptotic value is
  reached because now only the $H$ band is used to select stars.
\end{itemize}

\begin{figure}[t]
  \centering
  \includegraphics[width=\hsize,keepaspectratio]{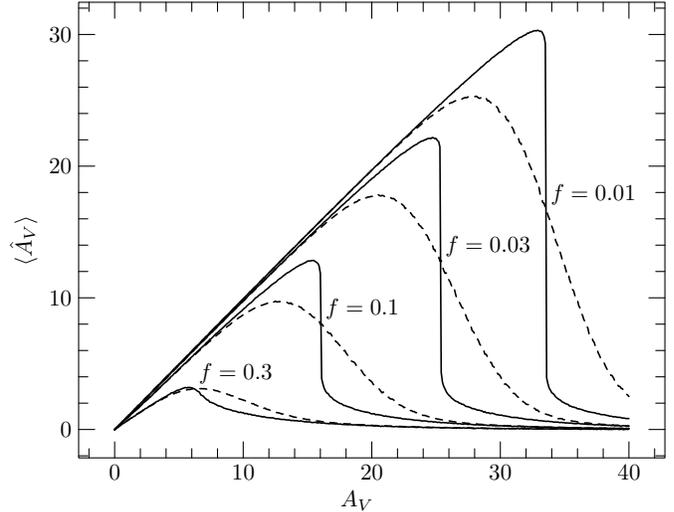}
  \caption{The bias for the \textsc{Nice} method due to the
    contamination by foreground stars.  Solid lines represents, for
    different fractions of foreground stars, the median of the
    distribution of the measured column densities.  Dashed lines,
    instead show the expectation value for the median of $\hat A_V$
    obtained from the measurements of 15 reddenings.}
  \label{fig:5}  
\end{figure}

\begin{figure}[t]
  \centering
  \includegraphics[width=\hsize,keepaspectratio]{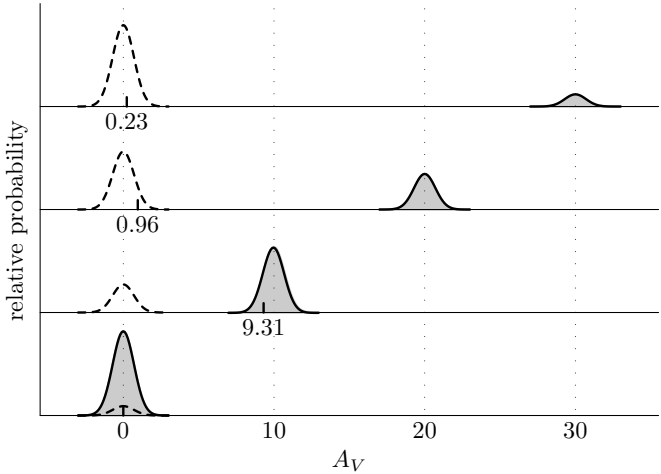}
  \caption{The effect of a foreground star contamination $f = 0.1$ on the
    distribution of column densities for different values of $A_V$.
    From the bottom to the top we show, for $A_V \in \{0,10,20,30\}$,
    both the distribution of background stars (solid plot, gray
    filled), and foreground stars (dashed plot).  Both distributions
    are taken to be Gaussian, the first centered on $A_V$, and the
    second on $0$.  As the column density increases, the relative
    contribution of foreground stars also increases because of
    selection effects.  At $A_V \simeq 16$, they become predominant
    and the median of the whole distribution quickly moves toward the
    foreground distribution.}
  \label{fig:6}
\end{figure}

A more serious problem is related to the contamination by foreground
stars, which can strongly bias our results in the direction of lower
column densities.  For regions with low extinction, where the expected
number of foreground stars is small compared to the expected number of
background stars, foreground star contamination is usually reduced by
using a \textit{median\/} estimator for $\hat A_V$, i.e.\ by averaging
the individual column densities measured in the direction of each star
with the median instead of the simple mean
\citep[e.g.][]{2002AJ....123.2559C, 2001A&A...377.1023L}.  As shown in
Fig.~\ref{fig:5}, this simple technique is very effective and leads to
\textit{almost\/} unbiased results at low column densities (see
Appendix~\ref{sec:medi-relat-estim} for a statistical discussion on
the median estimator).  However, for a given fraction of foreground
stars $f$ (which, as described above, is evaluated in regions with
negligible extinction) there is a threshold for $A_V$ after which the
cloud extinction makes the expected number of background stars smaller
than the expected number of foreground stars.  This threshold can be
evaluated using Eq.~\eqref{eq:43} and is
\begin{equation}
  \label{eq:53}
  A_V = \frac{1}{\alpha k_\lambda} \Log \frac{1-f}{f} \; .
\end{equation}
Because of the way the median estimator is defined, we observe in
Fig.~\ref{fig:5} an abrupt change on the measured column density $\hat
A_V$ close to this value (see also Fig.~\ref{fig:6}).  Note that the
relatively smooth transition observed in Fig.~\ref{fig:5} (solid
lines) is due to the intrinsic scatter in the star colors; indeed, in
absence of any scatter, we would observe an ``instantaneous'' change
from $\langle \hat A_V \rangle = A_V$ at low column densities to
$\langle \hat A_V \rangle = 0$ at high column densities.  In
Fig.~\ref{fig:5} we also plot the expectation value of a more
interesting quantity, the median over $N = 15$ measured column
densities.  This quantity differs from the median over the whole
distribution because of the statistical variations on the local number
of foreground stars.  In other words, since $N = 15$ is a relatively
small number, in different realizations of our simulations we can have
a significantly different number of foreground stars.  For example,
even at low column density, we can have a large fraction of foreground
stars; similarly, even at very large column density, there is a finite
probability to have the majority of stars in the background.  As a
result, the dashed line in Fig.~\ref{fig:5} has a much smoother
transition around the value given by Eq.~\eqref{eq:53}.  This is at
the same time a good and a bad news.  From one side, this means that
we can be significantly biased for values of $A_V$ smaller than the
threshold value of Eq.~\eqref{eq:53}; from the other side, this also
means that we can still partially make use of the median estimator at
relatively large column densities.

For regions with very high column density, it is normally quite easy
to identify and remove foreground stars because of their anomalous
colors with respect to the reddened background stars (see, e.g.,
\citealp{2001Natur.409..159A}).  Some authors
\citep[e.g.][]{2002AJ....123.2559C} make use of this information to
also remove stars in less dense regions using the following strategy.
The angular density $\sigma_\mathrm{f}$ of foreground stars is
determined using dense regions (where the foreground/background
identification is easy); then, for any region of the cloud, the $k$
bluer stars are thrown away, where $N_\mathrm{f} = \sigma_\mathrm{f}
\mathcal{A}$ is the expected number of foreground stars in the
analyzed region (deduced from the foreground density measured in the
dense regions).  This technique is quite simple and reasonably
effective, but unfortunately introduces a bias at low extinctions.
Consider, indeed, the limiting case where we have a negligible
extinction $A_V \simeq 0$.  The distinction between foreground and
background stars is in this situation ambiguous, and thus the
$N_\mathrm{f}$ bluer objects will likely include also some background
stars.  Hence, the results will be biased toward higher column
density.  The exact evaluation of the bias is non trivial, but can be
carried out using the techniques described in
App.~\ref{sec:medi-relat-estim}.  Here we report only an approximated
result valid for $A_V \simeq 0$ [see Eq.~\eqref{eq:80}],
\begin{equation}
  \label{eq:54}
  \langle \hat A_V \rangle \simeq \sqrt{2 \pi} \mathrm{Err}\bigl( \hat
  A_V \bigr) \frac{N_\mathrm{f}}{2 N} \; ,
\end{equation}
where $\mathrm{Err}\bigl( \hat A_V \bigr)$ is the average error of the
measured extinction from a single star.  Note that the result given in
Eq.~\eqref{eq:54} can be taken to be an upper limit for the bias,
since we expect this to decrease at high column densities, where the
identification of foreground stars is secure.

\section{A maximum-likelihood approach}
\label{sec:likelihood-approach}

From the discussion above, it is clear that both the star count and
the \textsc{Nice}(\textsc{r}) methods can produce unsatisfactory
results.  On one hand, the star count technique has a low
signal-to-noise ratio and produces significantly biased results at
high column densities (see Fig.~\ref{fig:2}).  On the other hand, the
color excess technique is more sensitive and has a smaller bias, but
can be severely affected by the contamination of foreground stars,
especially for large values of $A_V$.

As pointed out by \citet{2002AJ....123.2559C}, we can take advantage
of the different behavior of the star count and the color excess
techniques in the various column density regimes to partially solve
the problem of the contamination by foreground stars.  In particular,
\citeauthor{2002AJ....123.2559C} note that it is better to use the
color excess method at low column densities because of its higher
sensitivity, and to switch to the star count method at very large
column densities where the \textsc{Nice} method is completely
unreliable.  The optimal ``turning point'' where we need to change the
technique can be evaluated empirically by comparing the individual
results of the two methods at different column densities.

The solution proposed by \citeauthor{2002AJ....123.2559C} has the
significant advantage of being relatively simple to implement and
reasonably effective, but clearly is suboptimal in many respects:
\begin{itemize}
\item The choice of the ``turning point'' and of the matching strategy
  is to some extent arbitrary.
\item The overall estimate of $A_V$ still remains significantly biased
  at high column densities because at large $A_V$ the star counts are
  used (and this method has a non-negligible bias, see
  Fig.~\ref{fig:2}).
\item The density of foreground stars must still be determined
  separately and is taken to be constant on the whole field.
\end{itemize}

Using the theoretical framework developed so far, it is possible to
design and implement an efficient maximum-likelihood approach to the
problem.

\subsection{Likelihood}
\label{sec:likelihood}

Suppose again that in a region of the sky of area $\mathcal{A}$ we
observe in various bands $N$ stars with magnitudes $\{ \hat{\vec m}_n
\}$.  Assuming that the area of the sky is small enough so that there
are no significant changes on the relevant parameters of the problem
(such as the unreddened density $\sigma^{(0)}$, local expected density
$\sigma$, or the reddening probability distribution $p_{A_V}(A_V)$),
then we can easily evaluate the joint probability distribution for
such a star configuration.  First, we note that the number of stars
inside the region follows a Poisson distribution with average
$\mathcal{A} \sigma$:
\begin{equation}
  \label{eq:55}
  p_N(N) = \e^{-\mathcal{A} \sigma} \frac{(\mathcal{A}
  \sigma)^N}{N!} \; .
\end{equation}
The joint star probability distribution, i.e.\ the likelihood, is
given by
\begin{align}
  \label{eq:56}
  L \bigl( \{ \hat{\vec m}_n \} \bigr) = {} & p_N(N)
  \prod_{n=1}^N \frac{p_{\hat m}(\hat{\vec m}_n)}{\sigma} \notag\\
  {} = {} & \frac{\e^{-\mathcal{A} \sigma} A^N}{N!} \prod_{n=1}^N
  p_{\hat m}(\hat{\vec m}_n) \; .
\end{align}
Note that $L$ depends on unknown quantities, such as $p_{A_V}(A_V)$,
through $p_N(N)$, $p_{\hat m}(\hat{\vec m})$, and $\sigma$: hence,
assuming that there is no \textit{prior\/} for these unknown
quantities, we can obtain an estimate of them by maximizing $L$ or,
equivalently, $\ln L$.  In the following subsections we will
investigate in more detail this maximum-likelihood estimator.

\subsection{Implementation}
\label{sec:implementation}

We implemented the maximum-likelihood estimator using the
simplification described above.  In particular, we used the forms
\eqref{eq:34} and \eqref{eq:35} for the source magnitude probability
distribution $p_M(\vec M)$.  A simple calculation gives the following
relationships between the coefficients of Eq.~\eqref{eq:34} and
Eq.~\eqref{eq:35}:
\begin{align}
  \label{eq:57}
  P_{ab} = {} & C^{-1}_{ab} - \delta_{1a} \sum_{a'} C^{-1}_{a'b} -
  \delta_{1b} \sum_{b'} C^{-1}_{ab'} \notag\\
  & {} + \delta_{1a} \delta_{1b} \sum_{a',b'} C^{-1}_{a'b'} \; , \\
  \label{eq:58}
  Q_a = {} & {-\delta_{a1}} \alpha \ln 10 - C^{-1}_{ab} \chi_b +
  \delta_{1a} \sum_{a'} C^{-1}_{a'b} \chi_b \; , \\
  \label{eq:59}
  R = {} & {-2 \ln n} + C^{-1}_{ab} \chi_a \chi_b \; .
\end{align}

One of the advantages of the quadratic expression \eqref{eq:35} is
that we can write the effects of reddening as a simple transformation
of the three quantities $P$, $\vec Q$, and $R$.  In particular, the
transformation $\vec M \mapsto \vec M + \vec k A_V$ can be rewritten
as
\begin{align}
  \label{eq:60}
  P \mapsto {} & P \; , \qquad
  \vec Q \mapsto \vec Q - A_V P \vec k \notag\\
  R \mapsto {} & R - 2 A_V \vec Q^\mathrm{T} \vec k + A_V^2 \vec k^T P
  \vec k \; .
\end{align}
Hence, reddening does not change the functional form of the
probability distribution \eqref{eq:35} but only the three parameters
involved.

For the following discussion, it is also convenient to introduce a new
vector $\vec\mu \equiv (\vec M, 1) = (M_1, M_2, \dots, M_\Lambda, 1)$,
and a $(\Lambda+1) \times (\Lambda+1)$ matrix $S$ defined as
\begin{equation}
  \label{eq:61}
  S^{-1} = 
  \begin{pmatrix}
    P & \vec Q \\
    \vec Q^\mathrm{T} & R
  \end{pmatrix} 
  = 
  \begin{pmatrix}
    P_{11} & \cdots & P_{1\Lambda} & Q_1 \\
    \vdots & \ddots & \vdots & \vdots \\
    P_{\Lambda 1} & \cdots & P_{\Lambda \Lambda} & Q_\Lambda \\
    Q_1 & \dots & Q_\Lambda & R
  \end{pmatrix} \; .
\end{equation}
Then we can write the quadratic form appearing in Eq.~\eqref{eq:35}
simply as
\begin{equation}
  \label{eq:62}
  p_M(\vec M) = \exp \left[ - \frac{\vec\mu^\mathrm{T} S^{-1} \vec\mu}{2}
  \right] \; .
\end{equation}
More importantly, the action of measurement errors can be described as
simple transformations of $S$, provided that the measurement error on
each band can be described as a simple Gaussian with vanishing average
and variance $\vec v$.  In other words, assuming that the errors can
be represented as a convolution with the Gaussian kernel
\begin{equation}
  \label{eq:63}
  p_\epsilon(\vec\epsilon) = \frac{1}{(2\pi)^{\Lambda/2}
  \prod_{\lambda=1}^\Lambda v_\lambda} \exp\biggl[
  -\sum_{\lambda=1}^\Lambda \frac{\epsilon^2_\lambda}{2 v_\lambda^2 }
  \biggr] \; , 
\end{equation}
then the convolution $p_M * \Sigma$ can be written as
\begin{equation}
  \label{eq:64}
  (p_M * \Sigma)(\vec\mu) = \sqrt{\frac{\det S}{\det T}} \exp \left[
  -\frac{\vec\mu^\mathrm{T} T^{-1} \vec\mu}{2} \right] \; ,
\end{equation}
where $T = S + \mathrm{diag}(v_1^2, v_2^2, \dots, v_\Lambda^2, 0)$.

In conclusion, the simple implementation of the maximum-likelihood
method considered here can be summarized in the following items:
\begin{enumerate}
\item The quantities $P$, $\vec Q$, and $R$ appearing in
  Eq.~\eqref{eq:35} are determined in a control field free from any
  extinction.  Similarly, for each band the photometric errors, i.e.\ 
  the functions $p_{\epsilon \lambda}(\hat m_\lambda | m_\lambda)$,
  are calculated.
\item A cloud model (for example, the thin cloud approximation) is
  chosen, and the relevant parameters (for example, $A_V$ and $f$) are
  taken to be unknown and constant on the field.
\item For a given combination of the cloud parameters, the likelihood
  function \eqref{eq:56} is evaluated using the various observed
  magnitudes.  Note that the evaluation of the expected star density
  $\sigma$ and of the star probability $p_{\hat m}(\hat{\vec m})$ when
  the star is undetected in one or more bands is non-trivial and
  requires integrations over the completeness functions [cf.\ 
  Eq.~\eqref{eq:32}].
\item The last step is repeated for different values of the cloud
  parameters and the ones corresponding to the minimum of the
  likelihood function are taken as best estimates.
\item The local curvature of $L$ (i.e., the matrix of its second
  derivatives) is used to estimate the errors on the cloud parameters.
\end{enumerate}

Clearly, one key point here is the speed of the likelihood function,
which needs to be evaluated several times in our maximum-likelihood
approach.  In our implementation, the likelihood function has been
optimized by performing the relevant integrations (cf.\ point 3 above)
using appropriate bounds.  In other words, when an integration of
$p_{\hat m}(\hat{\vec m})$ was requested, we estimated the area in the
magnitude space where this function was significantly different from
zero, and performed the integral inside that area (as opposed to
performing the integral over the whole parameter space).  This
optimization was found to have a significant impact on the overall
speed of our implementation.

\subsection{Simulations}
\label{sec:simulations}

\begin{table}[b]
  \centering
  \begin{tabular}{cclc}
    Par. & Value & Description & ref.\ Eq. \\
    \hline
    $\alpha$ & $0.34$ & Slope of number counts & \eqref{eq:34} \\
    $\chi_1$ & $0.18$ & Average color $\langle H - K \rangle$ &
    \eqref{eq:34} \\
    $\chi_2$ & $0.82$ & Average color $\langle J - K \rangle$ &
    \eqref{eq:34} \\
    $C_{11}$ & $0.0078$ & Variance $\langle (H - K - \chi_1)^2 \rangle$ &
    \eqref{eq:34} \\
    $C_{12}$ & $0.0112$ & Covariance & \\[-2pt]
    & & $\langle (J - K - \chi_1) (H - K - \chi_2) \rangle$ &
    \eqref{eq:34} \\
    $C_{22}$ & $0.0375$ & Variance $\langle (J - K - \chi_2)^2 \rangle$ &
    \eqref{eq:34} \\
    $k_1$ & $0.112$ & Reddening law in $K$ band & \eqref{eq:8} \\
    $k_2$ & $0.175$ & Reddening law in $H$ band & \eqref{eq:8} \\
    $k_3$ & $0.282$ & Reddening law in $J$ band & \eqref{eq:8} \\
    $m^\mathrm{lim}_1$ & $14.3$ & Limit magnitude in $K$ band &
    \eqref{eq:20} \& \eqref{eq:27} \\
    $m^\mathrm{lim}_2$ & $14.9$ & Limit magnitude in $H$ band &
    \eqref{eq:20} \& \eqref{eq:27} \\
    $m^\mathrm{lim}_3$ & $15.8$ & Limit magnitude in $J$ band &
    \eqref{eq:20} \& \eqref{eq:27} \\
    $v_1$ & $0.05$ & Average photometric error \\[-2pt]
    & & $\langle (\hat m_1 - m_1)^2 \rangle^{1/2}$ on $K$ band &
    \eqref{eq:20} \\ 
    $v_2$ & $0.05$ & Average photometric error \\[-2pt]
    & & $\langle (\hat m_2 - m_2)^2 \rangle^{1/2}$ on $H$ band &
    \eqref{eq:20} \\ 
    $v_3$ & $0.05$ & Average photometric error \\[-2pt]
    & & $\langle (\hat m_3 - m_3)^2 \rangle^{1/2}$ on $J$ band &
    \eqref{eq:20}
  \end{tabular}
  \caption{The common parameters used for all numerical simulations.}
  \label{tab:2}
\end{table}

\begin{figure}[t]
  \centering
  \includegraphics[width=\hsize,keepaspectratio]{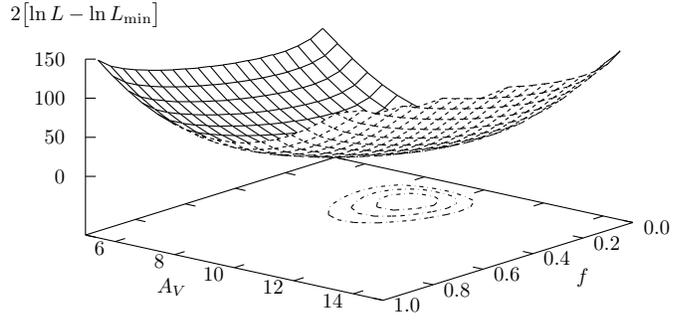}
  \caption{Log-likelihood surface plot.  The plot shows the
    logarithmic of the likelihood ratio as a function of the two
    unknown parameters $A_V$ and $f$ (the real values used for the
    simulation are $10$ and $0.2$ respectively).  The simulation has
    been carried out using $\sigma \mathcal{A} = 25$, but the
    actual number of stars generated were $N = 19$ (because of the
    Poisson noise on the number of stars).  On the bottom we also plot
    contours corresponding to $68.2\%$, $95.5\%$, and $99.7\%$
    confidence level regions.  Note that the likelihood is very smooth
    and has only a single well defined minimum.}
  \label{fig:7}
\end{figure}

The reliability and effectiveness of the maximum-likelihood approach
were assessed through extensive numerical simulations.  The
simulations were designed to reproduce with reasonable accuracy the
2MASS near-infrared data.  We simulated star observations in three
bands, $J$, $H$, and $K$, and used the various parameters described in
Table~\ref{tab:2}.

We initially considered an area of the sky $\mathcal{A}$ and a thin
cloud characterized by a fraction of foreground stars $f$ [cf.\ 
Eq.~\eqref{eq:18}] and a reddening $A_V$.  We randomly generated there
stars inside this area using the following recipe:
\begin{enumerate}
\item We evaluated the expected local star density $\sigma$ using
  Eqs.~\eqref{eq:31} and \eqref{eq:32}.  We found that the needed
  integrations could be performed more efficiently using a Monte-Carlo
  technique.
\item We calculated the effective number of stars $N$ by generating a
  random integer distributed according to a Poisson distribution with
  average $\mathcal{A} \sigma$.
\item For each of the $N$ star we adopted the following procedure:
  \begin{enumerate}
  \item We generated the unreddened magnitudes in the three bands
    according to Eq.~\eqref{eq:34}.
  \item We then uniformly generated a random number in the range $[0,
    1]$, and considered the star to be in the foreground if this
    number was smaller than $f$ [defined according to
    Eq.~\eqref{eq:18}].
  \item Background stars were reddened by adding $k_\lambda A_V$ to each
    magnitude; the magnitudes of foreground stars were left
    unchanged.
  \item We added photometric errors to all stars; these, for
    simplicity, were taken to be Gaussian distributed with standard
    deviation $v_\lambda = 0.05$ independent of the band and of the
    original magnitude.
  \item For each band, we uniformly generated a random number in the
    range $[0, 1]$, and took the star to be detected in the band if
    this number was smaller than the completeness function $\hat
    c_\lambda(\hat m_\lambda)$.  In the simulations discussed here we
    used for simplicity Heaviside functions for the completeness
    functions.
  \item We finally retained the star if it was detected in at least
    one band; otherwise, we repeated all steps above from point (a).
  \end{enumerate}
\end{enumerate}
In summary, at the end of a single star generation we had for each
star the three magnitudes in the bands $J$, $H$, and $K$ (with
possibly some magnitudes $\hat m_\lambda = \mathrm{null}$) and the
associated measurement errors $v_\lambda$.

\begin{figure}[t]
  \centering
  \includegraphics[width=\hsize,keepaspectratio]{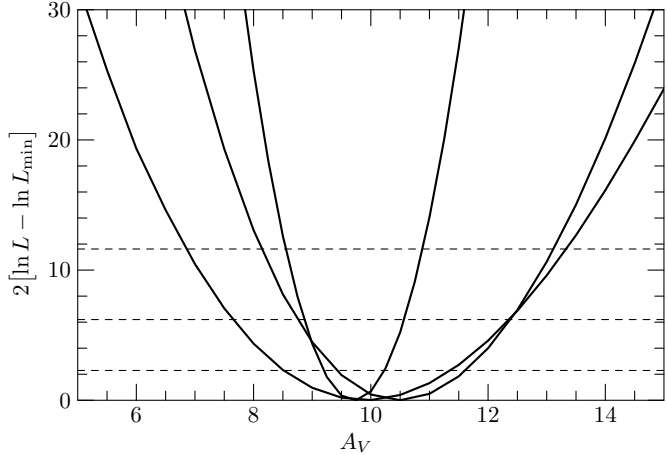}
  \caption{The log-likelihood (ratio) as a function of $A_V$ for different
    source densities, extremized for $f$.  The largest curve is
    obtained using $\sigma \mathcal{A} = 10$ (in this particular
    realization we had $N = 8$); the mid curve using
    $\sigma\mathcal{A} = 20$ ($N = 12$), and the most peaked curve
    using $\sigma\mathcal{A} = 40$ ($N = 35)$.  In all cases the
    true value of the column density was $A_V = 10$, and the fraction
    of foreground stars was $f = 0.2$.  Note that all curves are very
    well approximated by parabolas.  The intersections of the
    log-likelihood functions with the three dashed lines mark the
    $68.2\%$, $95.5\%$, and $99.7\%$ confidence level regions.}
  \label{fig:8}
\end{figure}

We then used this dataset to test the reliability and efficiency of
the maximum-likelihood estimator, and to compare it with the other
column density estimators considered in Sect.~\ref{sec:extinct-meas}.
The maximum-likelihood method was implemented as described in
Sect.~\ref{sec:implementation}, and was tested against the data
generated as described in the items above.

Figure~\ref{fig:7} shows the log-likelihood surface plot obtained in a
typical simulation.  The surface appears to be very smooth with a well
defined minimum, an essential condition for the reliability of the
maximum-likelihood approach.  Moreover, the log-likelihood function is
found to have approximately a parabolic shape, which further
simplifies the interpretation of the results obtained.  For example,
this property allows us to use the likelihood ratio \citep[see,
e.g.,][]{Eadie} to draw confidence level regions on the parameter
space (see the contours of Fig.~\ref{fig:7}).

Figure~\ref{fig:8} represents the log-likelihood as a function of
$A_V$ for three different datasets.  The figure was obtained by
minimizing the log-likelihood function with respect to $f$ for each
value of $A_V$ in the range $[5, 15]$, and by plotting the value of
this function.  

In order to assess more quantitatively the merits of the
maximum-likelihood method, we compared the statistical properties of
various column density estimators.  In particular, we simulated a
large number of ``observations'' using the technique described above,
and we studied the distribution of the column densities estimated
using the maximum-likelihood method, the \textsc{Nice}, and the
\textsc{Nicer} methods.  Simulations were carried out using a thin
cloud with true extinction in the range $A_V \in [0, 30]$ and with
different values of the foreground fraction $f$.  For the
\textsc{Nice} and \textsc{Nicer} estimators we used both the simple
mean and the median of the individual extinction measured for each
star; moreover, assuming that the density of foreground stars
$\sigma_\mathrm{f}$ could be determined separately, we discarded in
each simulation the $\sigma_\mathrm{f} \mathcal{A}$ bluer stars, and
used only the remaining (redder) stars in the analysis.  Note that in
some cases, for large values of $A_V$ and relatively large values of
$f$ we had no usable star for the \textsc{Nice} and \textsc{Nicer}
analysis; in other words, all stars left after the foreground
selection had only one band available (typically the $K$ band).  In
this case we just obtained a lower limit on $A_V$ by using the redder
usable star (even if this star was originally considered to be in the
foreground).

\begin{figure}[t!]
  \centering
  \includegraphics[width=\hsize,keepaspectratio]{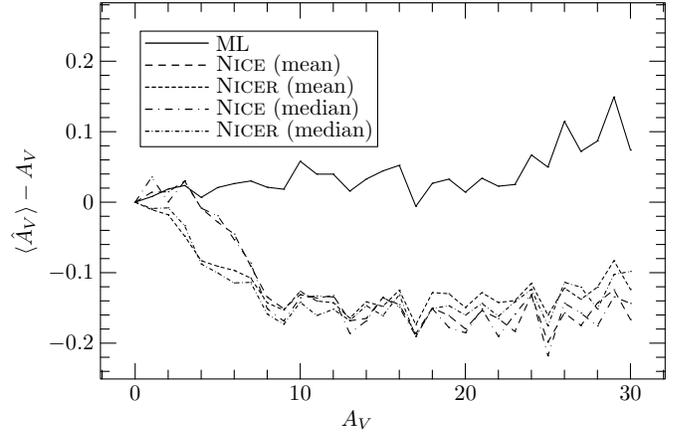}
  \caption{The bias $\langle \hat A_V \rangle - A_V$ of various column
    density measurement methods.  The bias is evaluated from the
    averages of $1\,000$ simulated fields with no foreground
    contribution ($f = 0$) and with average number of stars
    $\sigma\mathcal{A} = 25$.  The maximum-likelihood method,
    marked in the legend as ML for brevity, has negligible bias,
    especially for $A_V < 20$.}
  \label{fig:9}
\end{figure}

\begin{figure}[t!]
  \centering
  \includegraphics[width=\hsize,keepaspectratio]{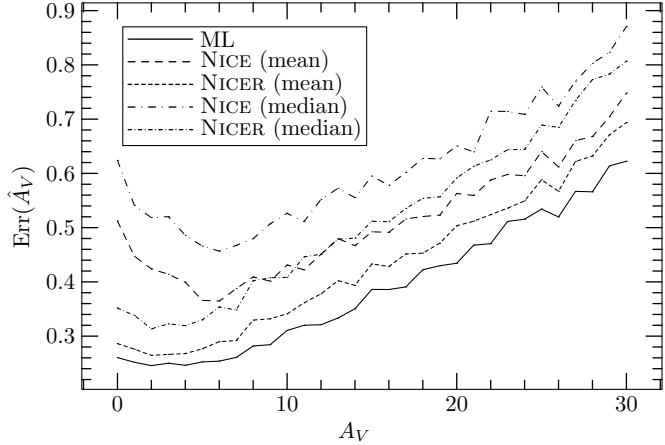}
  \caption{The total error $\bigl\langle \bigl( \hat A_V - A_V
    \bigr)^2 \bigr\rangle^{1/2}$ of various column density measurement
    methods, evaluated as in Fig.~\ref{fig:9}.}
  \label{fig:10}
\end{figure}

Figures~\ref{fig:9} and \ref{fig:10} show, respectively, the bias and
the total error obtained from the three methods considered here for
$A_V \in [0,30]$ and $f = 0$.  From these plots it is evident that the
maximum-likelihood estimator does not have any significant bias up to
$A_V = 20$ and a very small one (of the order of $0.1$) for larger
column densities.  Since for $f = 0$ we never have foreground stars,
the bias of the \textsc{Nice} and \textsc{Nicer} techniques does not
change if we use a mean or a median estimator.  Note also that the
bias in Fig.~\ref{fig:9} for these two methods is the one discussed in
detail above (cf.\ Fig.~\ref{fig:3}).  Regarding the total error, we
observe a steady increase of it for large values of $A_V$.  This can
be explained by noting that, although the average number of stars
$\sigma \mathcal{A} = 25$ is kept constant for all column densities,
when $A_V$ is large most stars are only detected in the $K$ band and
thus do not provide reddening information.  Figure~\ref{fig:10} also
shows that the maximum-likelihood method is clearly superior, although
$\textsc{Nicer}$ (with the mean estimator) also performs well.  As
expected, both median estimators are more noisy than the simple mean
(which, for $f = 0$, is optimal).

\begin{figure}[t!]
  \centering
  \includegraphics[width=\hsize,keepaspectratio]{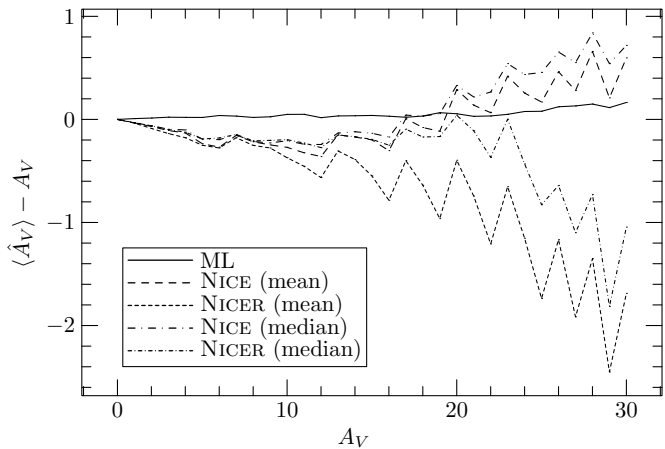}
  \caption{As for Fig.~\ref{fig:9}, but with $f = 0.02$.}
  \label{fig:11}
\end{figure}

\begin{figure}[t!]
  \centering
  \includegraphics[width=\hsize,keepaspectratio]{2267f12.eps}
  \caption{As for Fig.~\ref{fig:10}, but with $f = 0.02$.}
  \label{fig:12}
\end{figure}

\begin{figure}[t!]
  \centering
  \includegraphics[width=\hsize,keepaspectratio]{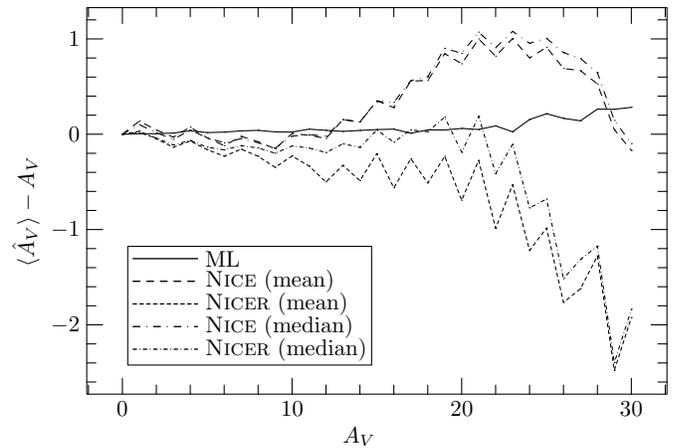}
  \caption{As for Fig.~\ref{fig:9}, but with $f = 0.05$.}
  \label{fig:13}
\end{figure}

\begin{figure}[t!]
  \centering
  \includegraphics[width=\hsize,keepaspectratio]{2267f14.eps}
  \caption{As for Fig.~\ref{fig:10}, but with $f = 0.05$.}
  \label{fig:14}
\end{figure}

\begin{figure}[t!]
  \centering
  \includegraphics[width=\hsize,keepaspectratio]{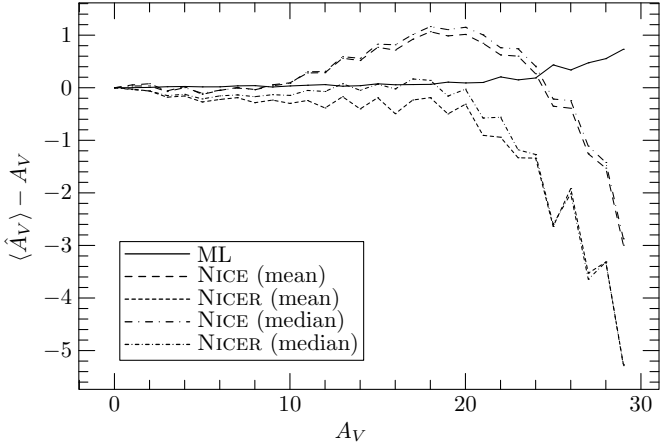}
  \caption{As for Fig.~\ref{fig:9}, but with $f = 0.10$.}
  \label{fig:15}
\end{figure}

\begin{figure}[t!]
  \centering
  \includegraphics[width=\hsize,keepaspectratio]{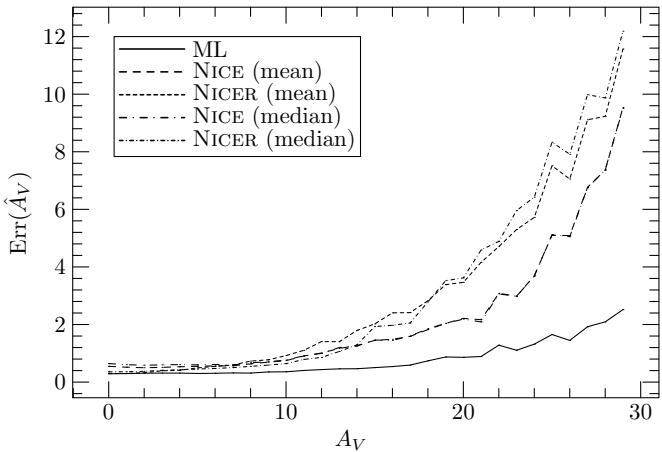}
  \caption{As for Fig.~\ref{fig:10}, but with $f = 0.10$.}
  \label{fig:16}
\end{figure}

The situation changes quite dramatically when $f > 0$.
Figures~\ref{fig:11}--\ref{fig:16} show the bias and the total error
of the various methods for increasing values of $f$.  A careful study
of these results can reveal several interesting characteristics of the
\textsc{Nice} and \textsc{Nicer} techniques.

Let us initially focus on the bias plots, shown in Figs.~\ref{fig:11},
\ref{fig:13}, and \ref{fig:15}.  At low column densities, i.e.\ for
$A_V < 10$, we find again the bias described in
Sect.~\ref{sec:near-infrared-color} and plotted in Fig.~\ref{fig:3}
(see also above Fig.~\ref{fig:9} for the case $f = 0$).  At median
column densities, i.e.\ for $A_V \sim 20$ (or $A_V \sim 30$ for $f =
0.02$), the bias can be explained by the selection effect due to the
correction of foreground stars.  As explained above [see
Eq.~\eqref{eq:54}], removing the $N_\mathrm{f} = \sigma_\mathrm{f}
\mathcal{A}$ bluer stars introduces a systematic error on the
extinction estimate.  This bias is positive (i.e.\ toward larger
extinctions) and can be as large as $1 \mbox{ mag}$ in $A_V$.
Finally, at high column densities ($A_V \sim 30$) both methods
systematically underestimate the extinction.  This is due to the heavy
contamination by foreground stars present at these large values of
reddening.  To better understand this point, we note that the
subtraction of the $N_\mathrm{f}$ bluer stars is a simplistic
approximation because the actual number of foreground stars is not
fixed (it is a Poisson random variable with average $N_\mathrm{f}$).
Depending on the number of foreground stars with respect to
$N_\mathrm{f}$ we can have three different situations:
\begin{itemize}
\item If the number of foreground stars is exactly $N_\mathrm{f}$,
  then at high column densities no bias is introduced and the estimate
  of $A_V$ is correctly performed;
\item For a simulation with a number of foreground stars smaller than
  $N_\mathrm{f}$, a small bias toward higher column densities is
  expected, because some of the bluer background stars are discarded;
\item Finally, if the number of foreground stars is underestimated, a
  very large bias toward smaller column densities is introduced.
\end{itemize}
Clearly, for large values of $A_V$ the third effect is expected to
dominate.  Indeed, Figs.~\ref{fig:13} and \ref{fig:15} show that both
the \textsc{Nice} and \textsc{Nicer} methods significantly
underestimate the reddening for large values of $A_V$.  In theory, as
mentioned above, when applying the \textsc{Nice} or \textsc{Nicer}
technique to high extinction regions, it should be relatively
straightforward to identify foreground stars by their color, and thus
the bias of these method could be smaller than suggested by the plot
shown here.  Note that apparently \textsc{Nicer} presents a larger
bias compared to \textsc{Nice}.  This is due to the larger flexibility
of the \textsc{Nicer} method, which is able to obtain a column density
estimate when any two of the three bands are available.  As a result,
\textsc{Nicer} is more affected by the contamination by foreground
stars described in the items above.  Indeed, our simulations show that
if we force \textsc{Nicer} to use only stars with observed $H$ and $K$
bands (i.e.\ essentially the same stars as the ones used by
\textsc{Nice}), its bias and its noise are drastically reduced and
become compatible with the ones of \textsc{Nice}.

Figures~\ref{fig:12}, \ref{fig:14}, and \ref{fig:16} show that the
total error of the \textsc{Nice} and \textsc{Nicer} methods increases
very rapidly at large column densities, where the contamination by
foreground stars is very likely; the maximum-likelihood estimator,
instead, has an almost flat error curve.  Hence, our novel method is
able to ``recognize'' the presence of foreground stars; moreover, the
inclusion of the background density in the likelihood expression
allows this estimator to ``switch'' to the number count technique in
regions with large extinction.  Figure~\ref{fig:12}, in particularly,
shows that even in extreme cases with a large foreground star
contamination the maximum-likelihood method is still very reliable and
accurate.  To better appreciate this point, we note that, in our
simulations, for $A_V = 25$ and $f = 0.2$ on average only one tenth of
the $\sim 25$ stars are background to the cloud.

\subsection{Limitations}
\label{sec:limitations}

The maximum-likelihood approach to the extinction measurements
presents clear advantages with respect to the standard techniques in
the simplified framework considered in this section (uniform $A_V$
over the cloud patch analysed, thin-cloud approximation, simple model
for the star intrinsic magnitude distribution).  One could thus
legitimately ask whether the maximum-likelihood technique is
applicable to more realistic and complex situations.

\subsubsection{Small-scale inhomogeneities}
\label{sec:small-scale-inhom}

Most clouds present clear sign of substructure at different scales and
a statistical analysis of the radio and NIR observations seems to
indicate the that these objects can be well described in terms of
turbulent models \cite[see, e.g.][]{1994ApJ...429..645M,
  1997ApJ...474..730P}.  In presence of significant inhomogeneities,
most methods (including the maximum-likelihood one) are expected to be
biased toward low values because the background stars will no longer
be randomly distributed in the patch of the sky used to estimate the
local extinction value, but will be preferentially detected in
low-extinction regions [cf.\ Eq.~\eqref{eq:43}].  Although a detailed
discussion of this effect is behind the scope of this paper, it is
worth considering the following points:
\begin{itemize}
\item For the $HK$ \textsc{Nice}, for the $JHK$ \textsc{Nicer}, and
  for the $JHK$ maximum-likelihood methods, small scale
  inhomogeneities become important when the local variations of $A_V$
  on the patch of the sky considered are of order $\Delta = 1 /
  (\alpha k_2 \ln 10)$, where $k_2 = A_H / A_V$ [cf.\ 
  Eq.~\eqref{eq:8}].  For typical 2MASS observations we find $\Delta
  \simeq 7.3 \mbox{ mag}$, and hence all methods should still be
  applicable to the analysis of regions with relatively low column
  densities (approximately $A_V < 10 \mbox{ mag}$).
\item The effect of substructures is studied in detail in a
  forthcoming paper \citep{L05}, where it is also presented a method
  to avoid the bias introduced by small-scale inhomogeneities.  The
  application this novel technique to a \textsc{Nicer} map of the Pipe
  nebula confirms the expectations summarized in the last item.  In
  particular, for the Pipe nebula the ``standard'' \textsc{Nicer} has
  a negligible small bias (below $0.2 \mbox{ mag}$) for $A_V < 10
  \mbox{ mag}$, while the bias increases dramatically for larger
  extinction values (e.g., it reaches $1 \mbox{ mag}$ at $A_V = 15
  \mbox{ mag}$).  A similar bias behavior is expected for the
  maximum-likelihood technique described in this section.
\item Since small-scale inhomogeneities are believed to be due to
  turbulent motions \citep{1981MNRAS.194..809L, 1997ApJ...474..730P,
    2004ApJ...615L..45H}, the probability distribution $p_{A_V}(A_V)$
  is expected to be a log-normal:
  \begin{equation}
    \label{eq:81}
    p_{A_V}(A_V) = \frac{1}{\sqrt{2 \pi} S A_V} \exp \biggl[ -
    \frac{(\ln A_V - M)^2}{2 S^2} \biggr] \; .
  \end{equation}
  Independently from the exact form of $p_{A_V}(A_V)$, the
  maximum-likelihood approach can be used in this more general
  framework to obtain the relevant parameters of $p_{A_V}$ (e.g., $M$
  and $S$ in the case of the log-normal of Eq.~\eqref{eq:81});
  moreover, the parameters estimated are asymptotically unbiased
  \citep[see][]{Eadie}.  We will consider the use of a non-trivial
  extinction probability distributions similar to the one of
  Eq.~\eqref{eq:81} in a follow-up paper.
\end{itemize}

\subsubsection{Star magnitude distribution}
\label{sec:star-magn-distr}

In the implementation of the maximum-likelihood technique discussed in
this section, we used the simple model for the magnitude probability
distribution $p_M(\vec M)$ [cf.\ Eq.~\eqref{eq:34}].  However, the
functional form of $p_M(\vec M)$ used can be inaccurate in describing
real data for several reasons:
\begin{itemize}
\item Different stellar populations can produce multiple peaks in the
  color distribution of stars.  For example, giant and dwarf stars
  produce two distinct peaks in the $J - H$ histogram (the two peaks
  are also clearly visible in a $JHK$ color-color diagram).
\item Different stellar populations can also have different slopes of
  the number counts, which could different significantly from the
  ``nominal'' value $\alpha \simeq 0.34$ \citep[see,
  e.g.][]{2002AJ....123.2559C}.  If the various stellar populations
  also have different intrinsic mean colors, then the simple model
  \eqref{eq:34} could lead to inaccurate extinction measurements.
\item The average NIR colors of stars in regions free of extinction
  are not completely independent of the star luminosity.  For example,
  a color-magnitude plot of 2MASS stars shows that the average $J - K$
  color slightly increases as the $K$ magnitude increases.  Since
  this effect is rather small, the associated bias is probably
  negligible in most cases; moreover, this effect can be included in
  the expression \eqref{eq:35} with a suitable choice of the
  coefficients $P$, $\vec Q$, and $R$.
\end{itemize}
All issues described above are strictly related to the simplified
description for the probability $p_M(\vec M)$ used here, and
\textit{not\/} to the maximum-likelihood method itself.  In other
words, it is possible (and relatively easy) to implement a
maximum-likelihood estimator based on a more realistic probability
distribution for the star magnitudes (e.g., synthetic stellar
population models; see \citealp{2004A&A...416..157R,
  1994ApJ...424..852J}).  For this purpose, we note that the most
computationally effective way to generalize $p_M(\vec M)$ is by
writing it as a linear combination of functions of the form
\eqref{eq:34} (with each function representing, \textit{de facto}, a
different star population).

\subsubsection{General remarks}
\label{sec:general-remarks}

On of most significant drawback of the maximum-likelihood approach is
its speed.  The implementation used in this paper is approximately one
order of magnitude slower than the \textsc{Nicer} method (at least on
a typical workstation), and this might prevent large applications of
the method proposed here.  On the other hand, the fast technological
progress in the computer speed justifies the work presented in this
paper, in the sense that soon it will be possible to use the
maximum-likelihood method on large datasets composed of millions of
stars.

Another possible issue related to the technique presented here is the
need for a more detailed knowledge of the properties of the data used.
As a comparison, we note that the original \textsc{Nice} technique
makes use only of the $H$ and $K$ magnitudes of stars and of the
average color $\langle H - K \rangle$ of stars in the control field.
In addition to these data, \textsc{Nicer} also requires the estimated
errors of star magnitudes in the NIR bands used.  Finally, the
maximum-likelihood method requires a detailed knowledge of the
probability distribution $p_M(\vec M)$, of the measurement errors
$p_{\epsilon_\lambda}(\hat m_\lambda | m_\lambda)$ of each star, and
of the completeness functions $\hat c_\lambda(\hat m_\lambda)$.  In
reality, in the Bayesian approach implicitly adopted in this paper,
the maximum-likelihood method can also be used if an
\textit{approximate\/} knowledge of these parameters is available.
Suppose, for example, that the parameters $P$, $\vec Q$, and $R$ that
characterize $p_M(\vec M)$ [see Eq.~\eqref{eq:35}] are only
approximately known from the measurements in the control field.  In
this case we can take these parameters as \textit{unknowns\/} in the
expression for the likelihood, and we can thus minimize $L$ with
respect to them as well as with respect to the parameters that
characterize $p_{A_V}(A_V)$ ($A_V$ and $f$ in the case considered
here).  As customary in standard maximum-likelihood problems, we
include the knowledge on $P$, $\vec Q$, and $R$ as a \textit{prior\/}
in the function $L$.  Note that this way the dimension of the space
over which we need to minimize $L$ greatly increases (and this can
pose severe computational problems), but in principle the schema
proposed is applicable to real cases.

Interestingly, the \textsc{Nice} and \textsc{Nicer} technique can be
seen as special cases of the maximum-likelihood method when no prior
knowledge on $R$ is available: the complete lack information on the
normalization of $p_M(\vec M)$ forces our method to use the only color
information of the stars.  Similarly, the number counts method can be
recovered as special case when the knowledge on the average colors
$\chi_a$ is absent [or, equivalently, when the scatter in the colors
$C_{ab}$ is very large; cf.\ Eq.~\eqref{eq:34}].  This suggests that
the \textsc{Nice}(\textsc{r})) techniques are not more robust of the
maximum-likelihood one, but just simpler.

Finally, we mention that the presence of young stellar objects (that
could be for example embedded in the dense cores) can in principle
affect the extinction measurements.  Hence, these objects should be
``manually'' removed before using any extinction measurement method,
including the maximum-likelihood described in this paper.

\section{Discussion and conclusions}
\label{sec:discussion}

In this paper we have considered the problem of an accurate
determination of the extinction toward a dark molecular cloud.  The
results obtained here can be summarized in the following items:
\begin{itemize}
\item The extinction and the reddening of background stars have been
  considered from a general statistical point of view.
\item The bias and uncertainties of the two main NIR techniques used
  to map the extinction, the star count and the color excess methods,
  have been discussed in detail.  We have shown that, although the
  color excess method has generally a smaller error, it is affected by
  a large bias in presence of contamination by foreground stars.  We
  have also shown that both \textsc{Nice} and \textsc{Nicer} are
  affected by a relatively small bias (approximately $0.2
  \mathrm{mag}$ in $A_\mathrm{V}$) as a result of selection effects.
\item A new optimal maximum-likelihood method has been presented and
  tested with extensive simulations.
\end{itemize}

The simulations described in Sect.~\ref{sec:simulations} have shown
that \textit{in the simple case of a thin cloud the maximum-likelihood
  estimator performs significantly better than the \textsc{Nice} and
  \textsc{Nicer} estimators\/} (since the number count method is known
to have a larger noise, we did not report a detailed comparison with
this method here).  However, the maximum-likelihood approach also
allows us to consider more general cloud configurations, which cannot
be easily dealt with using standard techniques.  For example, the
maximum-likelihood techniques could be used to measure directly on the
same patch of the sky both the column density $A_V$ and the fraction
of foreground stars $f$; alternatively, it would be also possible to
determine $f$ globally on the cloud and take it as a constant on the
whole field (a good approximation for small clouds).

In Sect.~\ref{sec:small-scale-inhom} we discussed one of the most
serious limitations of NIR color excess studies in molecular clouds,
namely the bias introduced by substructures.  In our original
formulation, the maximum-likelihood method can be used not only in the
thin cloud approximation, but in general for any functional form of
$p_{A_V}(A_V)$.  Hence, as mentioned above, we could implement the
maximum-likelihood method using a more realistic probability
distribution for $A_V$, such as the one of Eq.~\eqref{eq:81}.

Another possibility offered by the maximum-likelihood approach is the
generalization of the thin-cloud approximation to a multi-layer case,
where two or more (thin) clouds located at different distances are
observed on overlapping areas of the sky.  For example, in case of a
double cloud we could write [cf.\ Eq.~\eqref{eq:17}]
\begin{align}
  \label{eq:65} 
  p_{A_V}(A_V) = {} & f^{(1)} \delta(A_V) + f^{(2)} \delta(A_V -
  A_V^{(1)}) \notag\\
  & {} + \bigl( 1 - f^{(1)} - f^{(2)} \bigr) \delta \bigl( A_V -
  A_V^{(1)} - A_V^{(2)} \bigr)  \; .
\end{align}
This configuration, for example, might be appropriate to study clouds
close to the galactic center, where the superposition of different
complexes is very likely.  Such a method could effectively disentangle
the effects of the two clouds provided the values of $f^{(1)}$ and
$f^{(2)}$ are sufficiently different.  Alternatively, one could use a
double cloud as a null test, i.e. to check that indeed the thin-cloud
approximation is appropriate (in this case one expects to find
$A_V^{(1)} \simeq 0$, $A_V^{(2)} \simeq 0$, or $f^{(1)} \simeq 0$).
Some of these possibilities will be investigated in detail in a
follow-up paper by using 2MASS data.

\acknowledgements 

We are very grateful to G.~Bertin and J.~Alves for useful and
stimulating discussions, and to the referee, L.~Cambresy for providing
stimulating comments and for helping us improve the paper
significantly.

\appendix

\section{The median and related estimators}
\label{sec:medi-relat-estim}

The goal of this appendix is to derive the probability distribution of
the median of $n$ independent identical random variables.  This
analysis is useful to address some of the issues discussed in
Sect.~\ref{sec:near-infrared-color}.

\subsection{Notation}
\label{sec:notation}

In the following we will often deal with \textit{ordered\/} and
\textit{unordered\/} $n$-tuples.  The latter can be taken to be a set
of $n$ elements, where typically each element is a random variable or
an element of another tuple; the former can be taken to be a list of
$n$ elements, where thus each element is associated to a position in
the list.  Hence, given a positive integer $k \le n$, it makes sense
to talk about the $k$-th element of an ordered $n$-tuple, while this
is meaningless for an unordered tuple.

We will denote an ordered $n$-tuple using brackets, as in $[x_1,
\dots, x_n]$; instead, we will reserve braces for the unordered
tuples, as in $\{ x_1, \dots, x_n \}$.  Note that, for consistency
with the definition, we identify unordered tuples if these differ just
by a permutation of the elements: thus $\{ x_1, x_2, x_3 \}$
is, for instance, identical to $\{ x_2, x_1, x_3 \}$.

\subsection{$P^n_k$ and the median estimator}
\label{sec:pn_k-medi-estim}

Let us call $p(x)$ the probability distribution for the real random
variable $x$, and let us consider the generation of unordered
$n$-tuples $\{x_1, x_2, \dots, x_n \}$ of such variables.  Suppose
that, after the random generation of a tuple, we order the tuple such
that $x_1 \le x_2 \le \dots \le x_n$.  We consider now the probability
distribution $p^n_k(x_k)$ for the $k$-th element of the ordered tuple
$[x_1, x_2, \dots, x_n]$, which can be written as
\begin{equation}
  \label{eq:66}
  p^n_k(x_k) = n \binom{n-1}{k-1} p(x_k) \bigl[ P(x_k) \bigr]^{k-1}
  \bigl[ 1 - P(x_k) \bigr]^{n - k} \; ,
\end{equation}
where $P(x)$ is the cumulative probability distribution of $x$:
\begin{equation}
  \label{eq:67}
  P(x) = \int_{-\infty}^x p(x') \, \diff x' \; .
\end{equation}
The expression appearing in Eq.~\eqref{eq:66} can be explained as
follows.  Let us consider an element (for example the first) of the
unordered $n$-tuple.  The probability that this element is the $k$-th
in the ordered tuple and that it has a value in the range $[x, x +
\diff x]$ is the product of three terms:
\begin{itemize}
\item $p(x) \, \diff x$ to takes into account the intrinsic
  probability distribution of $x$;
\item $\binom{n-1}{k-1} \bigl[ P(x_k) \bigr]^{k-1} \bigl[ 1 - P(x_k)
  \bigr]^{n - k}$, which is a simple binomial distribution giving the
  probability that the value chosen has $k-1$ elements of the order
  $n$-tuple at its left, and $n - k$ elements at its right.
\item Finally, we have to multiply the whole result by $n$ in order to
  take into account the arbitrary choice of the $k$-th element in the
  $n$-tuple.
\end{itemize}
  
\begin{figure}[t]
  \centering
  \includegraphics[width=\hsize,keepaspectratio]{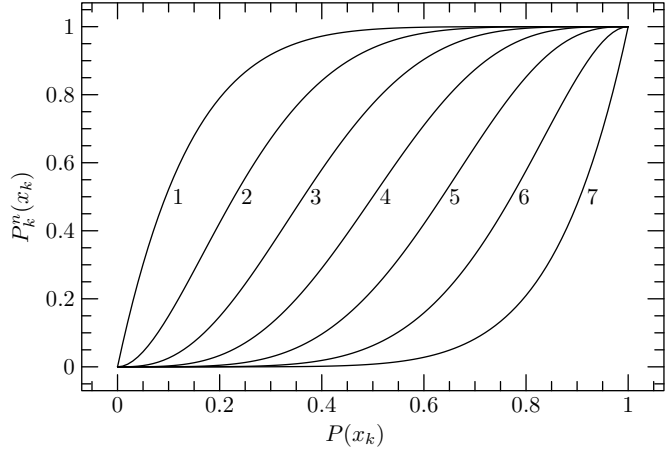}
  \caption{The cumulative probability distribution $P^n_k(x_k)$ as a
  function of $P(x)$ for $n=7$ and various values of $k$ (marked close
  to the relative curves).}
  \label{fig:A1}
\end{figure}

\begin{figure}[t]
  \centering
  \includegraphics[width=\hsize,keepaspectratio]{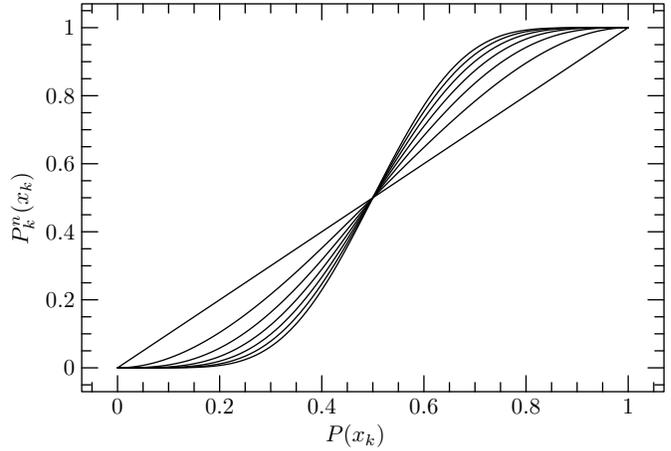}
  \caption{The cumulative probability distribution $P^n_k(x_k)$ as a
    function of $P(x)$ for $k \in \{1, 2, 3, 4, 5, 6, 7 \}$ and
    $n=2 k - 1$.  Note that all curves pass through the point $(0.5,
    0.5)$.}
  \label{fig:A2}
\end{figure}

The cumulative distribution $P^n_k(x_k)$ is given by
\begin{align}
  \label{eq:68}
  P^n_k(x_k) = {} & \int_{-\infty}^{x_k} p^n_k(x'_k) \, \diff x'_k
  \notag\\
  {} = {} & \frac{n!}{(k-1)! (n-k)!} \sum_{\ell=0}^{n-k} (-1)^\ell
  \binom{n-k}{\ell} \notag\\
  & {} \times \int_{-\infty}^{x_k} \bigl[P(x'_k) \bigr]^{k-1 +
    \ell} p(x'_k) \, \diff x'_k \notag\\
  {} = {} & \sum_{m=k}^{n} (-1)^{m + k} 
  \begin{pmatrix}
    n \\ m
  \end{pmatrix}
  \begin{pmatrix}
    m-1 \\ k-1
  \end{pmatrix}
  \bigl[ P(x_k) \bigr]^m \; .
\end{align}
In the last step we changed the index variable into $m = k + \ell$.
The final result obtained in Eq.~\eqref{eq:68} has the advantage
to be a simple (polynomial) expression in $P(x)$.

Figures~\ref{fig:A1} and \ref{fig:A2} show the polynomials
$P^n_k(x_k)$ as a function of $P(x_k)$ for various values of $n$ and
$k$.  These figures suggest a number of properties for $P^n_k(x_k)$
that can be verified analytically with the help of the equations
written above:
\begin{itemize}
\item $P^n_k(x) = 0$ if $P(x) = 0$, and $P^n_k(x) = 1$ if $P(x) = 1$;
  moreover, $P^n_k$ is a monotonic increasing function of $P$.  This
  implies that $p^n_k(x)$ vanishes where $p(x)$ vanishes.
\item Using Eq.~\eqref{eq:66}, it is possible to show that [see below
  Eq.~\eqref{eq:75}]
  \begin{align}
    \label{eq:69}
    \frac{1}{n} \sum_{k=1}^n P^n_k(x) = {} & P(x) \; , &
    \frac{1}{n} \sum_{k=1}^n p^n_k(x) = {} & p(x) \; .
  \end{align}
\item As suggested by Fig.~\ref{fig:A1} and by Eq.~\eqref{eq:66}, we
  have
  \begin{equation}
    \label{eq:70}
    P^n_k(x) = 1 - P^n_{n + 1 - k}(x') \; ,
  \end{equation}
  provided that $P(x) = 1 - P(x')$.  This, in particular, implies that
  $P^{2k-1}_k(x) = 1/2$ if $P(x) = 1/2$ (see Fig.~\ref{fig:A2}).
\item Equation~\eqref{eq:68} specialized to the cases $k = 1$ and $k =
  n$ is
  \begin{align}
    \label{eq:71}
    P^n_1(x_1) {} & = \bigl[ P(x_1) \bigr]^n \; , &
    P^n_n(x_n) {} & = 1 - \bigl[ 1 - P(x_n) \bigr]^n \; .
  \end{align}
\end{itemize}
The two last properties have a close relationship with the median
estimator.  We note, in fact, that for odd $n$ the median estimator is
in our notation $x_{(n+1)/2}$; as a result, the probability
distribution $p^{2k-1}_k(x_k)$ is just the probability distribution of
the median for $n = 2k-1$.  Hence, the property \eqref{eq:70} written
above can be rephrased as
\begin{quote}
  The median of the probability distribution of the median estimator
  of the random variable $x$ is the median of the probability
  distribution of $x$.
\end{quote}
Note that here we use the term ``median'' to denote both the usual
median of an $n$-tuple $\{ x_1, \dots, x_n \}$, and the median of a
distribution, defined as the value $x$ such that the value cumulative
distribution is $1/2$.

\subsection{$P^n_{\le k}$ and $P^n_{\ge k}$}
\label{sec:pn_k-pn_k}

For our purposes, it is also useful to define and study two
probability distributions closely related to $P^n_k(x_k)$.  Let us
consider again the ordered $n$-tuple $[ x_1, \dots, x_n ]$, where each
element is drawn from a probability distribution $p(x)$.  Now let us
retain only the first $k$ elements of the ordered $n$-tuple, and let
us call $p^n_{\le k}(x_{\le k})$ the probability distribution for each
element of the \textit{unordered\/} $k$-tuple $\{ x_1, \dots, x_k \}$;
similarly, we call $p^n_{\ge k}(x_{\ge k})$ the probability
distribution for the tuple $\{ x_k, \dots, x_n \}$ where only the last
$(n - k + 1)$ elements of the original ordered $n$-tuple are retained.

The evaluation of the probability distribution $p^n_{\ge k}(x)$ can be
broken into two parts.  Consider the $(k-1)$-th element $x_{k-1}$ of
the tuple $[ x_1, \dots, x_n ]$ (i.e., the last element discarded);
then, clearly, each element of the $(n-k+1)$-tuple $\{ x_k, \dots, x_n
\}$ cannot be smaller than $x_{k-1}$.  Moreover, each element of
this \textit{unordered\/} tuple is distributed between $x_{k-1}$
and $\infty$ according to the (truncated) original probability
distribution $p(x)$.  Finally, by repeating this argument for each
possible value of $x_{k-1}$ (weighted by $p^n_{k-1}$), we obtain the
expression
\begin{equation}
  \label{eq:72}
  p^n_{\ge k}(x) = \int_{-\infty}^x p^n_{k-1}(x_{k-1})
  \frac{p(x)}{1 - P(x_{k-1})} \, \diff x_{k-1} \; .
\end{equation}
Note that the term $1 - P(x_{k-1})$ is introduced here to correctly
normalize the truncated probability distribution $p(x)$.

As usual in this appendix, it is more convenient to consider the
cumulative probability distributions.  We can thus integrate $p^n_{\ge
  k}(x)$ over $x$ and obtain, after a few manipulations, a closed
expression for $P^n_{\ge k}(x)$.  Here, however, we prefer to follow a
different and simpler path.

Let us consider again the tuple $[ x_k, \dots, x_n ]$.  Each element
of this \textit{ordered\/} tuple follows the probability distribution
$p^n_k(x_k)$ discussed in the previous section.  Hence, if we consider
the \textit{unordered\/} tuple, the elements $\{ x_k, \dots, x_n \}$
will follow the \textit{average\/} distribution
\begin{equation}
  \label{eq:73}
  p^n_{\ge k}(x) = \frac{1}{n - k + 1} \sum_{k'=k}^n p^n_k(x) \; .
\end{equation}
By integrating both sides of this equation on $\diff x$, we can verify
that the same relation holds for the cumulative probability
distributions.  Let us then evaluate the sum
\begin{align}
  \label{eq:74}
  P^n_{\ge k}(x) = {} & \frac{1}{n - k + 1} \sum_{k'=k}^n P^n_k(x)
  \notag \\ 
  {} = {} & \frac{1}{n - k + 1} \sum_{m=k}^n  
  \begin{pmatrix}
    n \\
    m
  \end{pmatrix}
  \bigl[ P(x) \bigr]^m
  \notag\\
  & {} \times \sum_{k'=k}^m (-1)^{k+m}
  \begin{pmatrix}
    m-1 \\
    k'-1
  \end{pmatrix} \; .
\end{align}
We now consider two cases.  If $k = 1$, then the last sum can be taken
to be the binomial expansion of $(1 - 1)^{m-1}$, which vanishes for all
integers $m > 1$.  Hence we are left just with the case $m = 1$,
for which we obtain
\begin{equation}
  \label{eq:75}
  P^n_{\ge 1} = \frac{1}{n} \sum_{k'=k}^n P^n_k(x) = P(x) \; .
\end{equation}
This shows that each element of the \textit{unordered\/} $n$-tuple $\{
x_1, \dots, x_n \}$ follows the original distribution $p(x)$, a very
natural result indeed.

If, instead, $k > 1$, then using the properties of the binomial
coefficient we find
\begin{equation}
  \label{eq:76}
  P^n_{\ge k}(x) = \frac{1}{n - k + 1} \sum_{m=k}^n (-1)^{m + k} 
  \begin{pmatrix}
    n \\
    m
  \end{pmatrix}
  \begin{pmatrix}
    m - 2 \\
    k - 2
  \end{pmatrix}
  \bigl[ P(x) \bigr]^m \; .
\end{equation}
The similarity of this result with the last line of Eq.~\eqref{eq:68}
is rather surprising.

Similarly, we wish to investigate the cumulative probability
distribution $P^n_{\le k}(x)$ for the $k$-tuple $\{ x_1, \dots, x_k
\}$.  This quantity is better evaluated using $P^n_{\ge k}(x)$:
\begin{align}
  \label{eq:77}
  P^n_{\le k}(x) = {} & \frac{1}{k} \sum_{k'=1}^k P^n_{k'}(x) \notag\\
  {} = {} & \frac{1}{k} \biggl[ \sum_{k'=1}^n P^n_{k'}(x) -
  \sum_{k'=k+1}^n P^n_{k'}(x) \biggr] \notag\\
  {} = {} & \frac{1}{k} \biggl[ n P(x) - (n - k) P^n_{\ge k+1}(x)
  \biggr] \notag\\
  {} = {} & \frac{n}{k} P(x) + \frac{1}{k} \sum_{m=k+1}^n 
  \begin{pmatrix}
    n \\
    m
  \end{pmatrix}
  \begin{pmatrix}
    m-2 \\
    k-1
  \end{pmatrix}
  (-1)^{m + k} \notag\\
  & \phantom{\frac{n}{k} P(x) + \frac{1}{k} \sum_{m=k+1}^n}
  \times \bigl[ P(x) \bigr]^m \; .
\end{align}

% \begin{figure}[t]
%  \centering
%   \includegraphics[width=\hsize,keepaspectratio]{figure.103}
%   \caption{The derivative of $\diff P^n_k / \diff P$ evaluated at $P =
%     0.5$ for $n = 2 k - 1$.  This quantity is important in the
%     evaluation of @@@, and can be approximated very well by the
%     function $\sqrt{1.25 k - 0.25}$ (represented as a dashed line in
%     this plot) and, less accurately, by $\sqrt{k}$ (dotted line).}
%   \label{fig:A3}
% \end{figure}

As an application of the results obtained in this section, we evaluate
the bias introduced in the estimate of the column density when using
the technique described in Sect.~\ref{sec:near-infrared-color} to
remove foreground stars [see Eq.~\eqref{eq:54}].  Suppose that we
observe stars in a region with no significant extinction, so that both
foreground and background stars have the same distribution in colors.
For simplicity, we assume that the reddening estimates $\hat A_V$ for
each individual star is a Gaussian distribution with vanishing average
and variance $\sigma_{\hat A_V}$.  In this situation, if we exclude
the $k$ bluer stars (because they are taken to be foreground), we will
bias the estimate of $A_V$ toward large column densities.  In
particular, the column densities for the stars left will be
distributed according to $p^n_{\ge k+1}$.  The median of this
distribution can be evaluated as follows.  At $A_V = 0$ we have
$P(A_V) = 1/2$, while
\begin{equation}
  \label{eq:78}
  P^n_{\ge k+1}(0) = \frac{1}{n - k} \biggl[ \frac{n}{2} -
  \sum_{k'=1}^k P^n_{k'}(1/2) \biggr] \; .
\end{equation}
Assuming $n \gg k$, then each term in the sum above is approximately
unity (cf.\ above Fig.~\ref{fig:A1}).  Hence we obtain
\begin{equation}
  \label{eq:79}
  P^n_{\ge k+1}(0) \simeq \frac{n - 2k}{2n - 2k} \; .
\end{equation}
The fact that this quantity is not exactly $1/2$ is telling us that
there is a bias.  The exact amount of this bias can be evaluated by
solving the equation $P^n_{\ge k + 1}(x) = 1/2$.  Using Newton's
method we obtain then the approximate solution
\begin{equation}
  \label{eq:80}
  x \simeq \left[ \frac{\diff P^n_{\ge k+1}}{\diff x}\right]^{-1}_{x=0} \left[
  \frac{1}{2} - P^n_{\ge k+1}(0) \right] \simeq \frac{k}{2 n p(0)} \; .
\end{equation}
Note that in this equation we used the relationship between cumulative
and differential probability distributions; moreover, we simplified
$p^n_k(0) \simeq 0$ at low $k$, and retained only terms linear in
$k/n$.  The final result obtained is thus given by Eq.~\eqref{eq:54}.

\bibliographystyle{aa} 
\bibliography{../dark-refs.bib}

\end{document}